\documentclass[sigplan]{acmart} 

\def\BibTeX{{\rm B\kern-.05em{\sc i\kern-.025em b}\kern-.08emT\kern-.1667em\lower.7ex\hbox{E}\kern-.125emX}}
      
\copyrightyear{2021}
\acmYear{2021}
\setcopyright{acmlicensed}
\usepackage{verbatim}  
\usepackage{algorithm}
\usepackage{algpseudocode}  
\usepackage{subfigure}
\usepackage{multirow}
\usepackage{threeparttable}

\begin{document}

%
\title{An Efficient Vectorization Scheme for Stencil Computation}
\renewcommand{\shorttitle}{An Efficient Vectorization Scheme for Stencil Computation}
\author{Kun Li}
\affiliation{%
  \institution{Institute of Computing Technology, Chinese Academy of Sciences} 
  \institution{University of Chinese Academy of Sciences} 
  \city{Beijing}
  \state{China}}
\email{likungw@gmail.com}
\author{Liang Yuan}
\affiliation{%
  \institution{Institute of Computing Technology, Chinese Academy of Sciences} 
  \city{Beijing}
  \state{China}} 
\email{yuanliang@ict.ac.cn} 
\author{Yunquan Zhang} 
\affiliation{%
  \institution{Institute of Computing Technology, Chinese Academy of Sciences} 
  \city{Beijing}
  \state{China}} 
\email{zyq@ict.ac.cn} 
\author{Yue Yue}
\affiliation{%
  \institution{Institute of Computing Technology, Chinese Academy of Sciences} 
  \institution{University of Chinese Academy of Sciences} 
  \city{Beijing}
  \state{China}}
\email{yyue1998@gmail.com}
\author{Hang Cao}
\affiliation{%
  \institution{Institute of Computing Technology, Chinese Academy of Sciences} 
  \institution{University of Chinese Academy of Sciences} 
  \city{Beijing}
  \state{China}}
\email{caohang@ict.ac.cn}
\author{Pengqi Lu}
\affiliation{%
  \institution{Institute of Computing Technology, Chinese Academy of Sciences} 
  \institution{University of Chinese Academy of Sciences} 
  \city{Beijing}
  \state{China}}
\email{lupengqi18s@ict.ac.cn} 
\begin{abstract}
  Stencil computation is one of the most important kernels in various scientific and engineering applications. A variety of work has focused on vectorization and tiling techniques, aiming at exploiting the in-core data parallelism and data locality respectively. In this paper, the downsides of existing vectorization schemes are analyzed. Briefly, they either incur data alignment conflicts or hurt the data locality when integrated with tiling. Then we propose a novel transpose layout to preserve the data locality for tiling and reduce the data reorganization overhead for vectorization simultaneously. To further improve the data reuse at the register level, a time loop unroll-and-jam strategy is designed to perform multistep stencil computation along the time dimension. Experimental results on the AVX-2 and AVX-512 CPUs show that our approach obtains a competitive performance. 
\end{abstract} 

%
%

%

\maketitle

\section{Introduction} 

Stencil is one of the most important kernels widely used across a set of scientific and engineering applications. It is extensively involved in various domains from physical simulations to machine learning \cite{li2019openkmc}. Stencil is also included as one of the seven computational motifs presented in the Berkeley View \cite{asanovic2008parallel,10.1145/3126908.3126920} and arises as a principal class of floating-point kernels in high-performance computing. 

A stencil contains a pre-defined pattern that updates each point in a $d$-dimensional spatial grid iteratively along the time dimension. The stencil's order \cite{10.1145/3126908.3126920,zhao2019exploiting} defines the dependent relationship in a certain direction. If the order of a symmetric stencil in one dimension is $r$, the value of one point at time $t$ is a weighted sum of $(2r+1)$ points at the previous time \cite{10.1145/1989493.1989508}. The naive implementation for a $d$-dimensional stencil contains $d+1$ loops where the time dimension is traversed in the outmost loop and all grid points are updated in inner loops. Since stencil is characterized by this regular computational structure, it is inherently a bandwidth-bound kernel with a low arithmetic intensity and poor data reuse \cite{yuan2019tessellating,Krishnamoorthy+:pldi07}.


Performance optimizations of stencils have been exhaustively investigated in the literature~\cite{Dursun+:jsupercomputing12,Luo+:ics15,Dutsch+:wosc14}. Traditional approaches have mainly focused on either vectorization or tiling schemes, aiming at improving the in-core data parallelism and the data locality in cache respectively. These two approaches are often regarded as two orthogonal methods working at different levels. Vectorization seeks to utilize the SIMD facilities in CPU to perform multiple data processing in parallel, while tiling tries to increase the reuse of a small set of data fit in cache. They actually complement each other and can be subtly combined. 


Prior work on vectorization of stencil computations primarily falls into two categories.
The first one is based on the associativity of the weighted sums of neighboring points. Specifically, the execution order of one stencil computation can be rearranged to exploit common sub-expression or data reuse at register or cache level 
\cite{zhao2019exploiting,7161520,rawat2018register,8665800}. 
Consequently, the number of load/store operations can be reduced and the bandwidth usage is alleviated in optimized execution order. 
The second one attempts to deal with the data alignment conflicts
\cite{henretty2011data,Henretty+:ics13},
which is the main performance-limiting factor.
The data alignment conflict is a problem caused by vectorization.
One milestone approach to address the data alignment conflict is the DLT method (Dimension-Lifting Transpose) \cite{henretty2011data}.
We will present a deep discussion on them in the next section.
  

As one of the crucial transformation techniques to exploit the parallelization and data locality for stencils, tiling, also known as blocking, has been widely studied for decades. Since the size of working sets in stencil-based applications is generally larger than the cache capacity on a processor, the spatial tiling algorithms are proposed to explore the data reuse by changing the traversal pattern of grid points
in one time step. Generally, a grid point in cache is utilized to perform stencil computation for all its neighbors before swapped out cache. Thus, the data transfers between the cache and main memory could be reduced. However, the improvement of such tiling techniques is restricted to the size of the neighbor pattern \cite{10.1145/3126908.3126920,Krishnamoorthy+:pldi07}. Temporal tiling techniques have been developed to allow more in-cache data reuse across the time dimension. 


The aforementioned two approaches of stencil computation optimizations
often have no influence on the implementation of each other.
The fundamental reason is that the vectorization typically
applies to the innermost loop.
Therefore, integrating one technique of vectorization with
another tiling scheme is often straightforward.
However, the data organization overhead for vectorization 
may degrade the data locality.
Furthermore, to the best of our knowledge, most of the prior work only focuses on temporal tiling technique on the cache level.
This only optimizes the data transfer volume between cache and memory
and the high bandwidth demands of CPU-cache communication is still
unaddressed or even worse with vectorization.
We will present a deep discussion of these two problems in the next section.

In this paper,  we first design a novel transpose layout to overcome the input data alignment conflicts of vectorization. 
The new layout is formed with an improved in-CPU matrix transpose scheme,
which achieves the lower bounds both on the total number of data organization operations
and the whole latency.
Compared with conventional methods, the corresponding computation scheme for the new layout requires 
less data organization operations,
whose cost can be further overlapped by arithmetic calculations.
To enhance the data reuse on the register level,
we propose an approach to perform multiple time steps for stencil computations.
The in-register data can be reused to perform successive updates along the time dimension, which has not explored in existing work. 
Finally, we integrated the proposed layout with
a tiling framework.
It only requires a slight modification of the new vectorization scheme
to preserve the data reuse ability of tiling.
The proposed vectorization scheme is evaluated with AVX-2 and AVX-512 instructions for 1D, 2D, and 3D stencils. The results show that our approach is obviously competitive with the existing highly-optimized work\cite{Henretty+:ics13,henretty2011data,10.1145/3126908.3126920}.
 


This paper makes the following contributions:
\begin{itemize}
\item We propose an efficient transpose layout
and corresponding vectorization scheme for stencil computation.
The layout transformation utilizes an improved matrix transpose
of the lowest latency.

\item We exploit the in-register data reuse by performing multiple time step computation based upon the new proposed transpose layout.

\item An integrated approach is proposed to perform a tiling framework in conjunction with the vectorization scheme. 

\item We demonstrate that the proposed approach could achieve superior performance compared to several highly-optimized stencil benchmarks on multi-core processors.

\end{itemize}

The paper is organized as follows. Section 2 presents the relevant background and elaborates on the addressed problem. Section 3 introduces the proposed vectorization scheme and the tiling technique. 
Section 4 provides experimental results that demonstrate our approach produces a higher performance compared to the benchmarks. In Section 5, we present the related work and Section 6 concludes the paper.

\section{Background}

\subsection{Data Alignment Conflicts of Vectorization}

\begin{figure}[t] 
  \begin{center}
  \centering
  \includegraphics[width=0.48\textwidth]{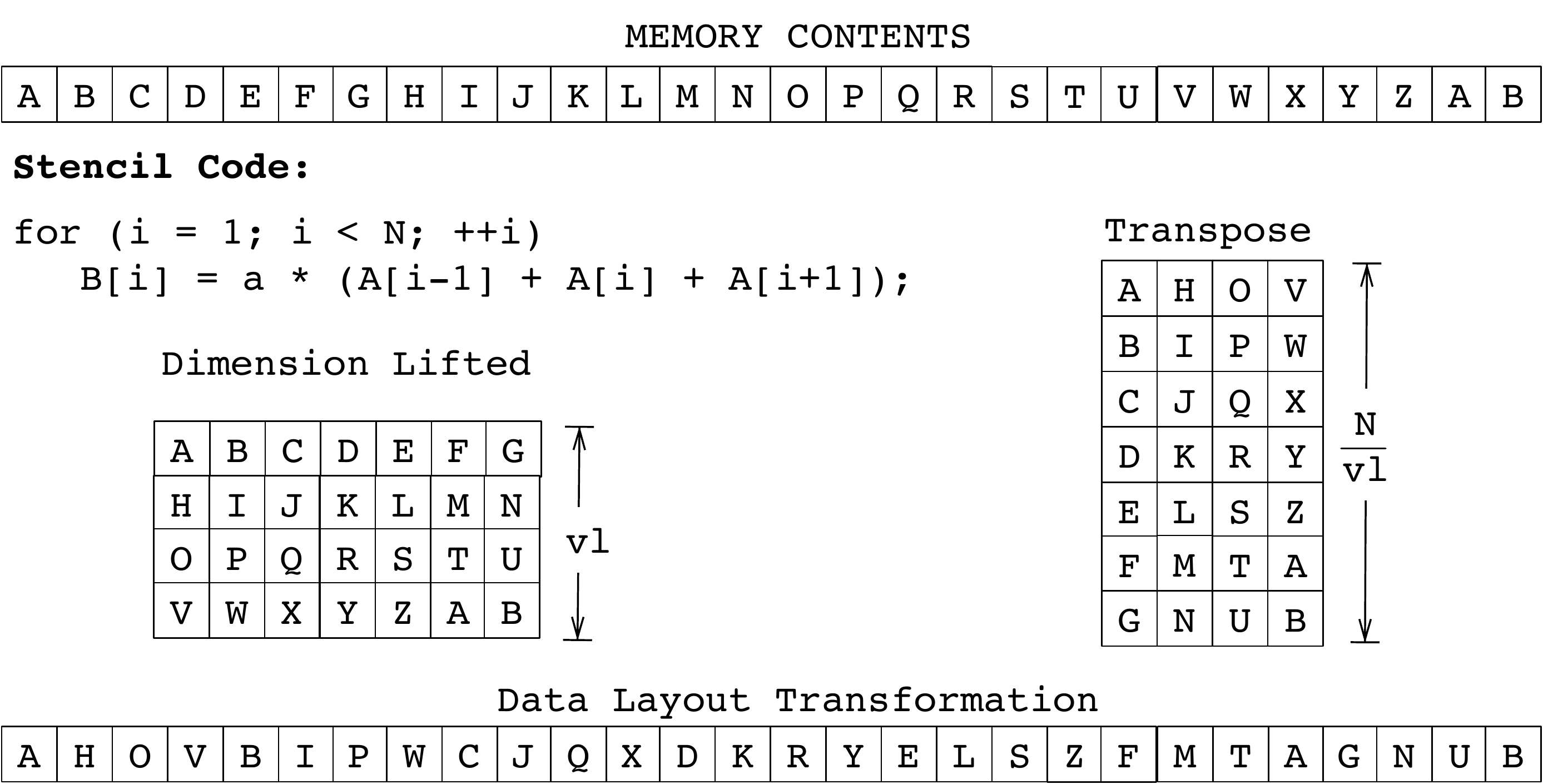}
  \caption{\label{bg}Illustration of input data alignment conflicts handling in DLT.}
  \end{center}
\end{figure}

We take the 1D3P stencil as an example to illustrate the fundamental problem of stencil computations caused by vectorization. Since in most existing work vectorization is restricted to innermost loops \cite{datta2009optimization}, the codes shown in Figure \ref{bg} only illustrates the 1D3P stencil of one time step.

In the $i$-th iteration of this scalar code execution, it loads $A[i+1]$ and $B[i]$ to registers and reuses register data $A[i-1]$ and $A[i]$ referenced by the previous calculation of $B[i-1]$. Observing the CPU-memory data transfer, this code is exactly similar to a common array copy code, i.e. the \verb=memcpy= function \cite{falke2013extending}. The computation implementation inside CPU is straightforward. Loop optimizations like loop unrolling also preserve these properties.

The vectorization groups a set of data in a vector register and processes them in parallel.
The naive vectorization of the 1D3P stencil code computes contiguous elements in the output array $B$. Assume the vector register holds 4 elements (vector length $vl=4$), the vectorization code performs the calculation using vector operations and output $(B[1],B[2],B[3],B[4])$ with one vector register.

A well-known problem incurred by the vectorization of stencil codes is the input data alignment conflicts. For example, to compute $(B[1],B[2],B[3],B[4])$, it requires three vectors: $(A[0],A[1],A[2],A[3])$, $(A[1],A[2],A[3],A[4])$ and $(A[2],A[3],A[4],A[5])$. The element $A[2]$ appears in all these vector registers but at different positions. We call this a data alignment conflict.  Thus there is no corresponding simple execution as the scalar code.

To address the data alignment problem, two common implementations are often adopted. The first one loads all the needed elements from memory in a vector form straightforward. Due to the low operational intensity, the stencil computation is often regarded as a memory-starving application. Compared with the scalar code, this multiple load vectorization method further increases the data transfer volume. Moreover, in each iteration of this code, it has at least two unaligned memory references where the first data address is not at a 32-byte boundary. Since CPU implementations favor aligned data loads and stores, these unaligned memory references will degrade the performance considerably.

The second solution is similar to the scalar code in terms of the CPU-memory data transfer. It loads each input element to vector register only once and assembles the required vectors via inter-register data permutations instructions. Compared with the multiple load method, this data permutations method reduces the memory bandwidth usage and takes the advantages of the rich set of data-reordering instructions supported by most SIMD architectures. However, the execution unit for data permutations inside the CPU may become the bottleneck.

\subsection{Dimension-Lifting Transpose (DLT)}

One milestone approach to address the data alignment conflict is the DLT method \cite{henretty2011data}. {In DLT the original one-dimensional array of length $N$ is viewed as a matrix of size $vl$*$(N/vl)$, where $vl$ is the vector length in vector elements. For example, $vl$=4 for double-precision floats in a 256-bit vector.} It then performs a global transpose. Figure \ref{bg} illustrates the DLT method for a one-dimensional array of 28 elements. The DLT layout overcomes the input data alignment conflicts. For instance, the second $vl=4$ elements in the transformed layout are formed into one vector (B[1], B[8], B[15],B[22] and all the three required input vectors: left vector (A[0], A[7], A[14], A[21]), center vector (A[1],A[8],A[15],A[23]) and right vector (A[2], A[9], A[15], A[23]) are free of data sharing and stored contiguous in memory. DLT needs to assemble input vectors for calculating output vectors at boundary.

DLT has the following disadvantages. First, DLT can be viewed as $vl$ independent stencils if we ignore the boundary processing. Therefore when incorporated with blocking frameworks, the data reuse decreases $vl$ times. The reason is that there is no data reuse among the $vl$ independent stencils. Second, DLT suffers from the overhead of explicit transpose operations executed before and after the stencil computation. For 1D and 2D stencils in scientific applications, the number of time loops is often large enough to amortize the transpose overhead.
But for 3D stencils and low-dimensional in other applications like image processing, the time size is small that makes the transpose overhead unignorable.
Finally, it's hard to implement the DLT transpose in-place and it often chooses to use an additional array to store the transposed data. This increases the space complexity of the code.

\section{The Transpose Layout}

In this section, we first discuss the drawbacks of existing vectorization methods.
Then 
we present a new transpose layout and its corresponding stencil computation scheme.
Next, we present several further optimizations on the transpose layout including
the extension to multiple time steps, integration with a tiling framework and
an improved matrix transpose algorithm. 

\subsection{Motivation}

From the hardware perspective,
the critical approach to boost performance is to fully utilize the execution units that perform the arithmetic instructions.
Since there is no data dependence in one time-step iteration of stencil computations, the only bottleneck is data preparation.
Equivalently, the key technique of vectorization
is to address the data alignment conflict.

Our starting point is the observation of the disadvantages of existing methods.
The DLT is a promising method that extremely reduces the data reorganization operations.
However, essentially the DLT vectorization format hurts the locality properties as mentioned above. In particular, the elements in one vector are distant,
thus there is no data reuse among them. 

On the contrary, the straightforward multiple load and data reorganization methods load contiguous element in one vector. They lead to the optimal data locality when integrated with a temporal tiling scheme.

These two methods seem to be at two extreme ends of a balance between the number of reorganization operations of data in CPU and the reuse ability of data in cache.
Our scheme seeks to preserve the data locality property
and employs the fundamental idea of DLT to improve the overhead of
data preparation.

\begin{figure}[t] 
  \begin{center}
  \centering
  \includegraphics[width=0.4\textwidth]{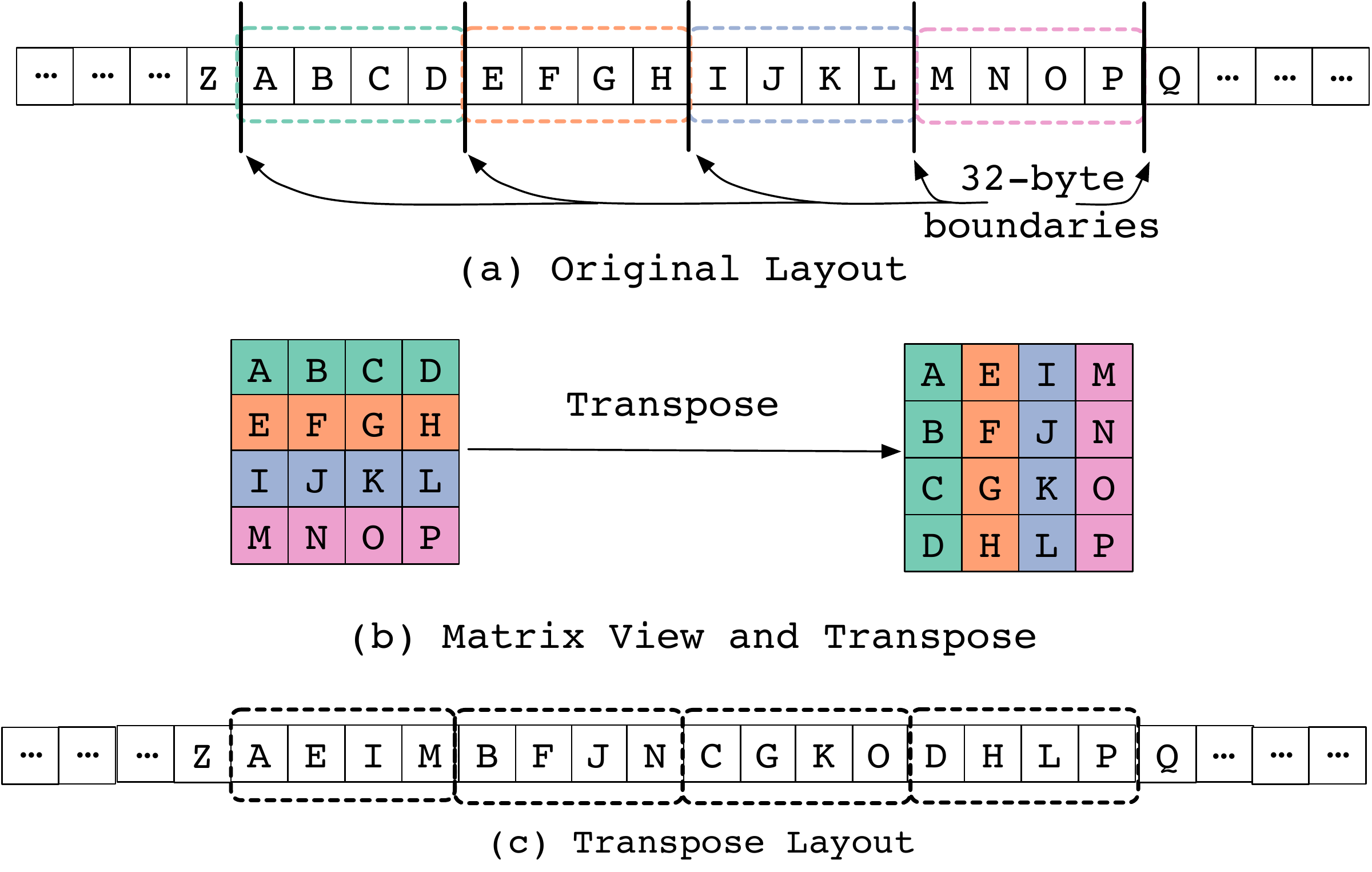}
  \caption{\label{rls}Register Transpose Layout for SIMD vector length of 4.}
  \end{center}
\end{figure}

\begin{figure} 
  \begin{center}
  \centering
  \includegraphics[width=0.3\textwidth]{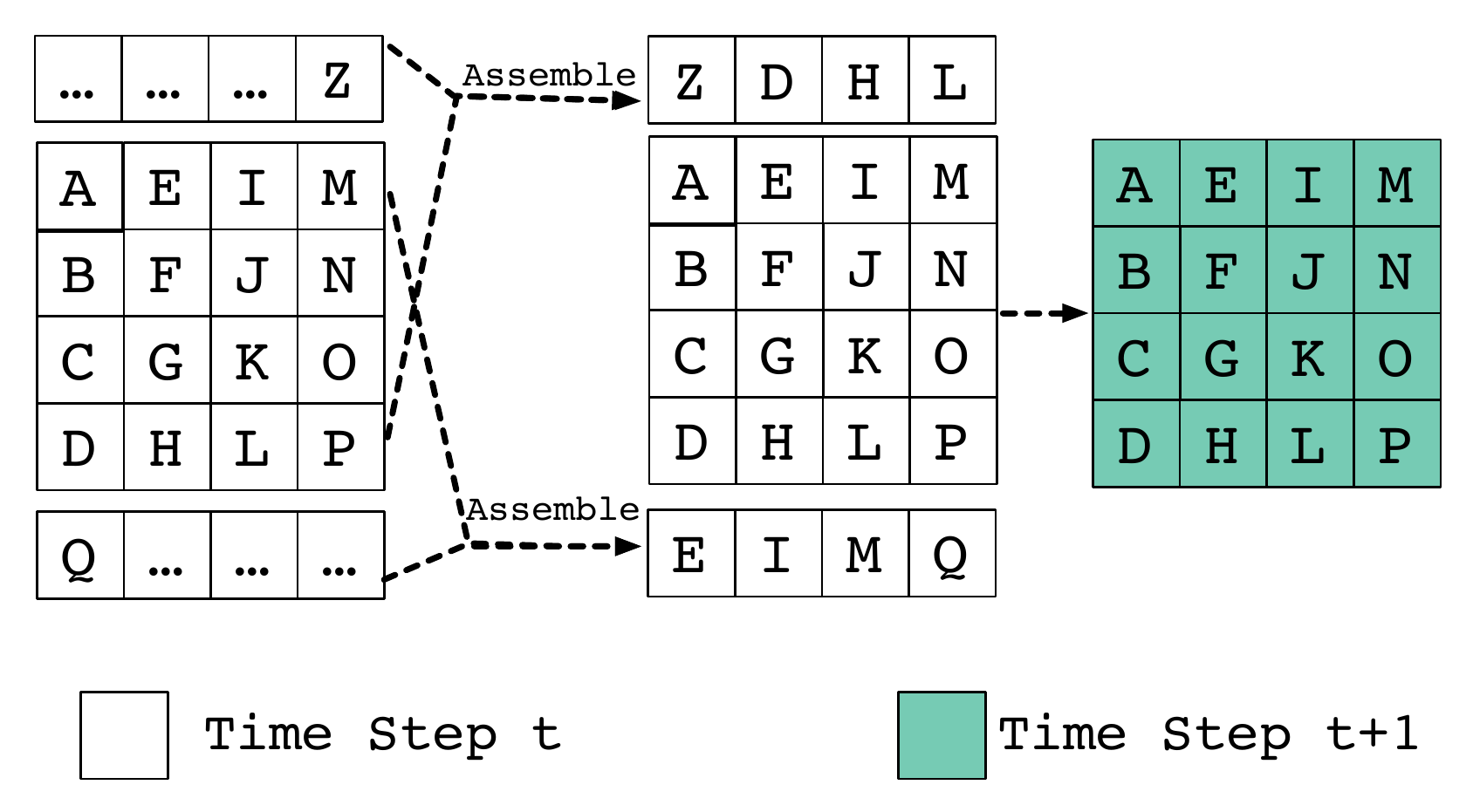}
  \caption{\label{multi} Illustration of stencil computation for transpose layout.}
  \end{center}
\end{figure}

\subsection{\label{subrls}Locally Transpose}

To preserve the data locality and 
reduce the number of data organization operations,
we apply a matrix transpose to a small sub-sequence of contiguous elements.
Specifically, like the dimension-lifting approach in DLT, 
the one-dimensional view to the sub-sequence is substituted by a two-dimensional
matrix view. To perform vectorization after a matrix transpose,
the column size of the matrix should be equal to the vector length $vl$.
Let the row size be $m$, the size of the matrix is then $vl*m$. 
Our locally transpose layout is equivalent to the DLT when $m=N/vl$
and the original data layout when $m=1$.

After the matrix transpose, it still requires some data reorganizations
for computing the first and last one of the $m$ vectors.
For example, if $m=1$, the original data preparations of the left and right vectors
must be done for computing each output vector.
This is the trade-off between data locality and the number of data preparations
explained above.

There are several considerations for deciding the size $m$.
First, $m$ should be large enough to hidden
the overhead of the data reorganizations for the first and last vectors
by the actually arithmetic operations of the middle vectors. Assume the order of a stencil is $r$, then the number of arithmetic operations of the middle vectors 
is $(2r+1)*(m-1)+1$. The number of data operations is $4r$
since the first and last vectors need $2r$ vectors and assembling
each of them requires two reorganization instructions as will 
be explained shortly.
Thus $m$ should be at least $3$.
Notice that this limitation is irrelevant to the order $r$.
Second, to avoid an additional array that is needed to store the transposed data as
in the DLT format, it's desirable to complete the matrix transpose in CPU.
Thus the $m$ input vectors and additional auxiliary vectors must be kept
in the CPU vector register file.
In this work we always set $m=vl$.
The final reason is that transposing a matrix of size $vl*vl$ is easier to implement on modern CPU products.
We will present a highly efficient algorithm
for matrix transpose of size $vl*vl$ later.


Figure \ref{rls} illustrates the transpose layout for a one-dimensional stencil with a vector length of four. The matrix transposes of every sub-sequence of $vl*vl$ length is performed before and after the stencil computation.
In the rest of the paper, we also refer to the $vl$ vectors as a \textbf{vector set} (VS).
Note that in the implementation a vector set is always aligned to a 32-Byte boundary.


The update of one vector set of the 1D3P stencil requires two assembled vectors.
One is the left dependent vector of its first vector
and the other is the right dependent vector of its last vector.
Figure \ref{multi} describes the data reorganization of these two vectors.
The first vector is $(A,E,I,M)$ and its left dependent vector is $(Z,D,H,L)$ which is stored in 
two distant vectors in the transpose layout, $(*,*,*,Z)$ and $(D,H,L,*)$.
These two vectors are combined by a blend instruction
followed by a permute operation to shift the components to the right circularly.

The stencil computations of the vector set are straightforward
as shown in Figure \ref{multi}.
We then achieve an efficient vectorization scheme by performing lower-overhead matrix transpose and two data operations per vector set. Moreover, the proposed vectorization scheme avoids data reloads compared with the multiple load method and frequent inter-vector permutations compared with the data reorganization method. The transpose layout could also be applied to higher-order and multidimensional stencils in the same manner.

\subsection{\label{FastNBL}Unroll-and-jam the Time Loop}

\begin{figure} 
  \begin{center}
  \centering
  \includegraphics[width=0.5\textwidth]{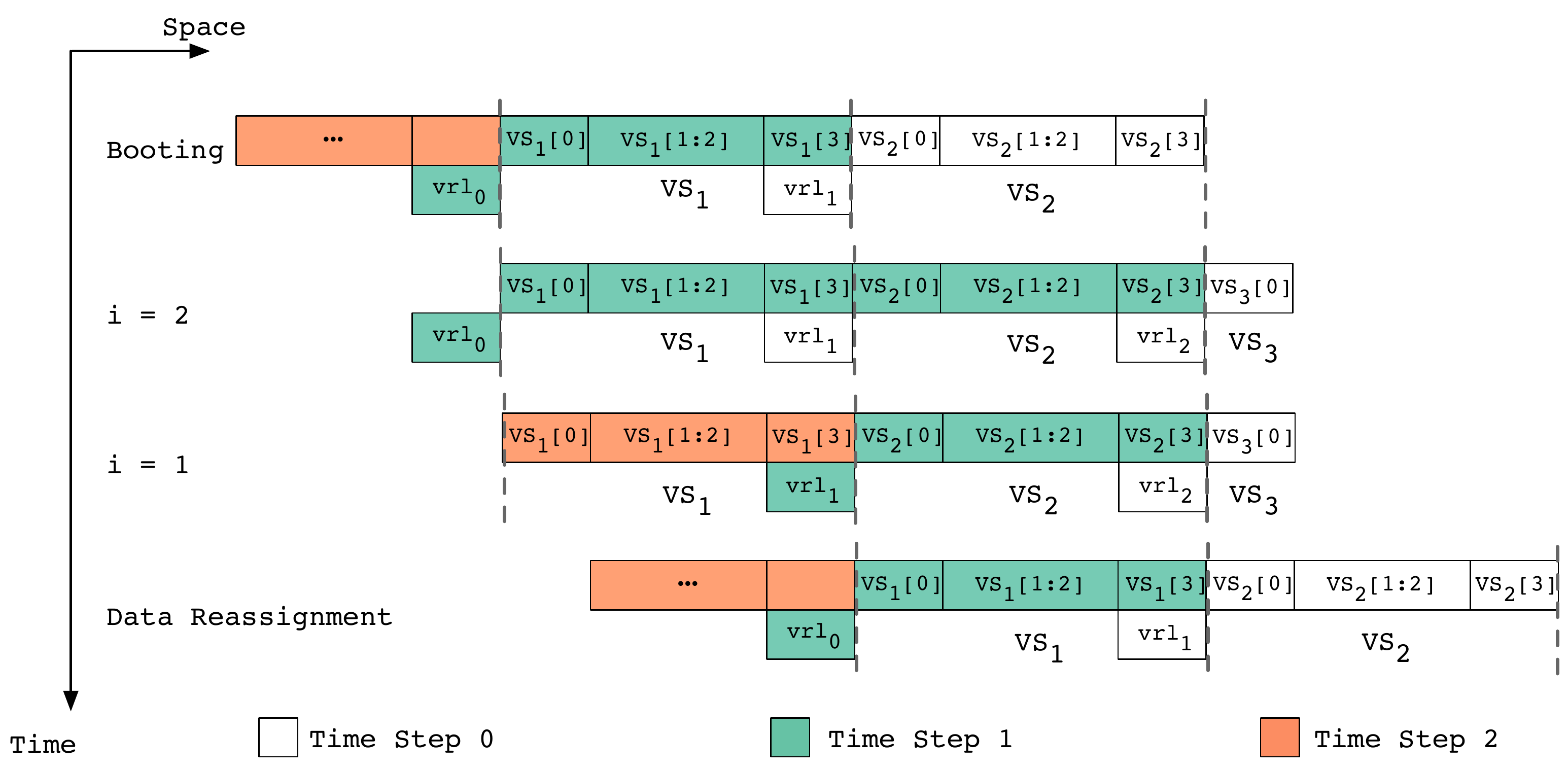}
  \caption{\label{multistep2} Illustration of stencil computation for two time steps.}
  \end{center}
\end{figure}

In general, stencil computation is restricted to its input data alignment conflicts, and all elements are only updated once before the round starts in the next time step. Although blocking technique \cite{yuan2019tessellating,10.1145/3126908.3126920,bandishti2012tiling} can be utilized to decrease the data transfers between main memory and cache, there is no in-register data reuse between successive time loops. 
Therefore the in-CPU flops/byte ratio is limited by the stencil pattern. To the best of our knowledge, computation for multiple time steps in registers is not explored in existing work. 

We develop an unroll-and-jam strategy of the time loop. 
It loads one element at time dimension $t$ and updates it $k$ time steps before store it to memory.
$k$ is called the unrolling factor.
The normal execution corresponds to the case of $k=1$.
If $k>1$ the execution is equivalent to unroll the time loop $k$ times and jam them.
Consequently, it improves the in-CPU flops/byte ratio by $k$ times. A one-dimensional example is illustrated in this subsection and the strategy is also applicable to multidimensional stencils.

\begin{algorithm}[t]  
  \caption{\label{algrls}Unroll-and-jam the Time Loop}  
  \begin{algorithmic}[1] 
      \Function {Assemble}{$\mathbf{v}_{a}$,$\mathbf{v}_{b}$}
      \State $\mathbf{v}_c$=\_mm256\_blend\_pd($\mathbf{v}_{a}$,$\mathbf{v}_b$)
      \State $\mathbf{v}_c$=\_mm256\_permute4$\times$64\_pd($\mathbf{v}_c$)
      \State \Return{$\mathbf{v}_c$}  
      \EndFunction
     \Function {Compute}{$\mathbf{v}_{left}$, $\mathbf{v}_1$, $\mathbf{v}_2$, $\mathbf{v}_3$, $\mathbf{v}_4$, $\mathbf{v}_{right}$} 
       \State $\mathbf{v}_0\gets$ \Call{Assemble}{$\mathbf{v}_{left}$,$\mathbf{v}_4$}
       \State $\mathbf{v}_5\gets$ \Call{Assemble}{$\mathbf{v}_{1}$,$\mathbf{v}_{right}$}
          \For{$i = 1 \to 4$}
              \State $\mathbf{v}_{i-1}\gets$ \Call{Stencil}{$\mathbf{v}_{i-1}$, $\mathbf{v}_i$, $\mathbf{v}_{i+1}$}
          \EndFor
  \State $\mathbf{v}_{1},\mathbf{v}_{2},\mathbf{v}_{3},\mathbf{v}_{4}\gets\mathbf{v}_{0},\mathbf{v}_{1},\mathbf{v}_{2},\mathbf{v}_{3}$
      \EndFunction  

      \Function{MultipleTimeSteps}{$\mathbf{VS}_{1:k}$,$\mathbf{vrl}_{0:k-1}$,$k$}
          \For{$j = k+1 \to N$} 
        \State $\mathbf{VS}_{k+1}\gets$ Load the $j$-th Vector Set
          
          
          \For{$i = k \to 1$}
           \State $\mathbf{vrl}_{i}\gets\mathbf{VS}_{i}[3]$ 
              \State \Call{Compute}{$\mathbf{vrl}_{i-1},\mathbf{VS}_{i}[0:3],\mathbf{VS}_{i+1}[0]$} 
          \EndFor
          \State Store $\mathbf{VS}_{1}$
        \For{$i = 1 \to k$}
        \State $\mathbf{VS}_{i}\gets\mathbf{VS}_{i+1}$ 
     \State  $\mathbf{vrl}_{i-1}\gets\mathbf{vrl}_{i}$ 
           \EndFor
          \EndFor   
           
      \EndFunction

      \Function{Main}{\ }
          \While{$t<T$}
              \State Booting computation.
              \State \Call{MultipleTimeSteps}{$\mathbf{VS}_{1:k}$,$\mathbf{vrl}_{0:k-1}$,$k$}
                          
              \State Epilogue computation.
              \State $t+=k$
          \EndWhile
      \EndFunction  

  \end{algorithmic}  
\end{algorithm}

Overall the algorithm is straightforward.
After update one vector set, we keep the result in registers
and process the next neighbor vector set.
Then the current vector set can be forwarded along time dimension one more step using the new value of the right neighbor.

Algorithm \ref{algrls} shows the pseudo-code of our multiple time steps updating scheme.
The C{\scshape ompute} function receives a set of $vl$ vectors and their dependent vectors that are assembled by the A{\scshape ssemble} function.
It computes the elements in the vector set by one time step.
Notice that this is an in-place updating that the value of last time will be overwritten.

The main function traverses the time loop stepped by the unrolling factor $k$.
For simplicity, we assume $T$ is divisible by $k$.
In each iteration of the while loop,
every element is forwarded $k$ steps along the time dimension.
The booting computation prepares the data at head needed by the 
following pipelined updating.
The top part of Figure \ref{multistep2} illustrates the case of $k=2$ after a booting computation.
The vector sets $\mathbf{VS}_1$ to $\mathbf{VS}_k$ have been updated $k-1$ to $0$ times, respectively.
Due to the overwriting property of the  C{\scshape ompute} function,
it needs to preserve the value of the last time of the vector to each vector set's left,
denoted as $\mathbf{vrl}_i$.
As the figure shows, $\mathbf{vrl}_i$ and $\mathbf{VS}_i[3]$ store the value of the same vector
at time $t-1$ and $t$, respectively.

The M{\scshape ultiple}T{\scshape ime}S{\scshape teps} function forwards all the vector sets from right to left by one time step.
Meanwhile, it preserves the old value of their rightmost vector in $\mathbf{vrl}$.
At the end of each iteration, $\mathbf{VS}_1$ has been updated $k$ times and is stored in memory.
Then after some data reassignments, the next loop is ready to execute.
Each iteration loads and stores one vector set of $vl*vl$ elements and performs 
$k*vl*vl$ stencil computations.
As mentioned above, it increases the in-CPU flops/byte ratio by $k$ times.

From the algorithm, we see that it needs $k$ vector sets and $k$ additional vectors to 
unroll-and-jam the time loop,
i.e., total $(vl+1)*k$ registers in addition to coefficient vector registers.
In modern CPUs, the typical number of available vector registers is $vl*4$, where
$vl$ is the capacity of double precision variables in one register,
therefore in this work we always set $k=2$.

There is another advantage of the algorithm.
Conventionally the stencil of Jacobi style 
is implemented with two arrays, storing
the value at odd and even time respectively.
If we set $k=2$, then the input and output value
are all at the even time. It's legal to reuse the input data space
and make the whole computation in-place.
The space usage is then reduced.

\subsection{Integrated With Tiling}

Vectorization and tiling are two orthogonal methods.
They target at different levels.
Vectorization boosts the computation using the data parallelism at the execution level, while tiling serves to exploit the data reuse at cache levels.
The transpose layout described above identifies a vectorization technique
as the solution to the data alignment conflict for stencils.
The multiple time update further improves the data reuse ability at the CPU vector register level.
In the following, we present the combination of the transpose layout and a tiling framework.



\begin{figure}[t]
  \label{cases}
  \begin{center}
  \centering
  \includegraphics[width=0.5\textwidth]{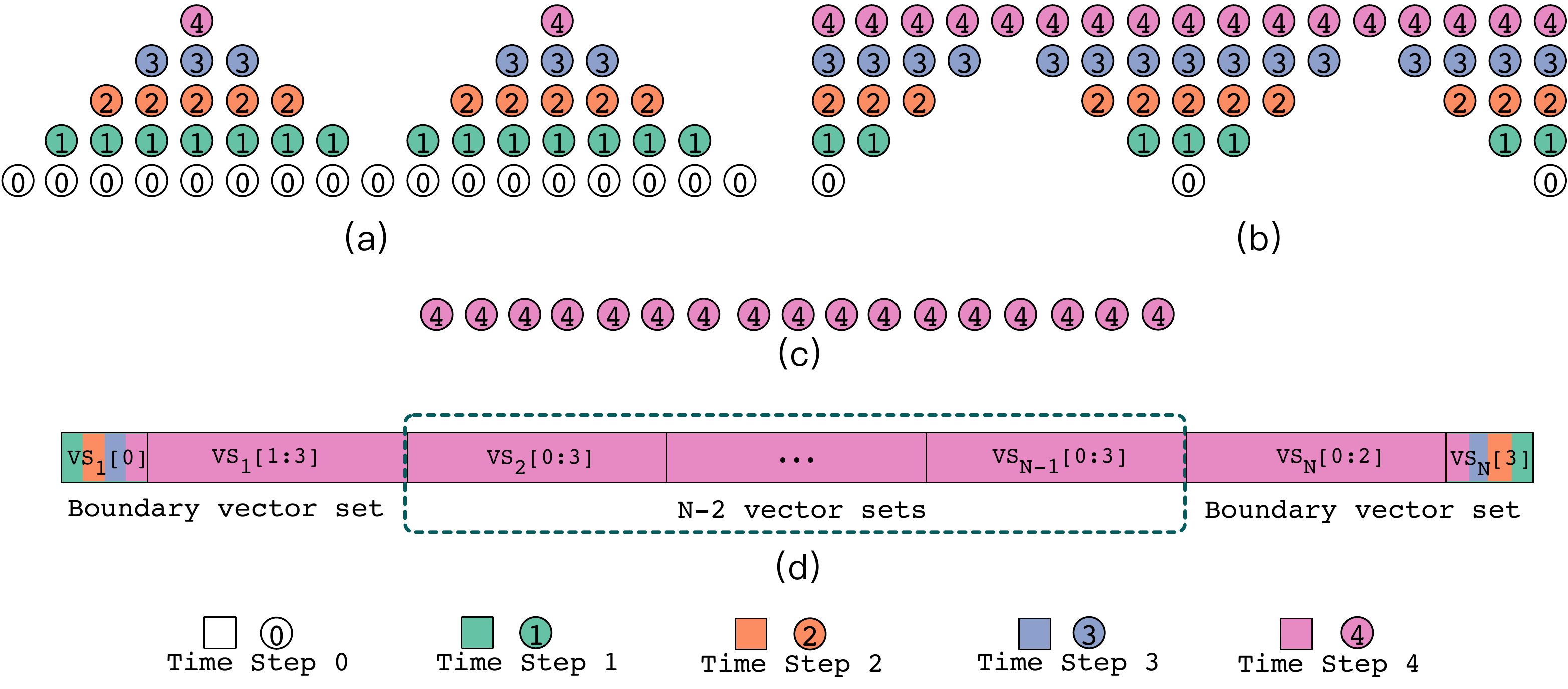}
  \caption{\label{tiles}Tessellate tiling iteration space for 1D updated with two time steps on register transpose layout.}
  \end{center}
  \vspace{-0.6cm}
\end{figure}

The tessellation tiling \cite{yuan2019tessellating} 
can be viewed as a tessellation in iteration space by utilizing shaped tiles.  Figure \ref{tiles} (a) and Figure \ref{tiles} (b) illustrate the tiling framework for a one-dimensional stencil. The iteration space is tessellated by triangles and inverted triangles in alternative stages. Thus, concurrent execution is processed by two stages which are started in each triangle with a given time range first, followed closely by the execution of inverted triangles over the same time range concurrently. 
Updates in different time steps are distinguished from each other by different colors, and the state of each element along the time dimension is represented with a number in Figure \ref{tiles}. For the example in Figure \ref{tiles} (a), the new state of each triangle contains (0,1,2,3,4,3,2,1,0) where the center element is updated four steps and its neighbors are updated fewer steps proportional to the distance with the center element. To make all elements updated with the same steps, two half parts from adjacent triangles constitute new inverted triangles and the elements are updated with the state (4,3,2,1,0,1,2,3,4). As Figure \ref{tiles} (c) shows, all elements are updated to four steps by adding the projection of the triangles with inverted triangles.
With the tessellate tiling strategy, concurrent execution for different tiles is enabled over a given time range without redundant computation. 

The only problem for applying the transpose layout is the calculations at the two boundaries of each block.
The execution of triangles is a 'shrinking' process, the range of processed elements decreases as the time forwards.
Similarly, an 'expanding' process occurs in the execution of inverted triangles.
Since the physical neighbor elements are stored apart from each other in one vector set,
the calculations of the vector set that covers a boundary are too complex to implement.
As the basic computing unit in the transpose layout is a vector set,
we convert the vector set at boundary back to the original format before the computation and 
employ a simple data reorganization method to process them. As illustrated in Figure \ref{tiles} (d), the shrinking and expanding process could be simplified in this way.
When the boundary slides away, the vector set is transposed again.



Further, the register transpose layout and time loop fusion make it feasible to achieve multiple time steps computation in registers over the tiles efficiently without reloading operations.

The tessellate tiling could also be applied for multidimensional stencil computations. For a $d$-dimensional stencil, tessellation in iteration space contains $d+1$ stages. Similar to the 1D stencil example in Figure \ref{tiles}, the spatial space in stage $i$ is tessellated by $tiles_i$ ($0\le i\le d$). $tiles_1$ is a hypercube (typically a line segment in 1D, square in 2D, cube in 3D). $tiles_{i+1}$ is built by recombining the sub-tiles split from adjacent $tiles_i$ along some dimensions.
Applying the transpose layout to higher-dimensional stencils is exactly similar to the one-dimensional case since the layout only
affects the unit-stride dimension.

\subsection{Transpose} 
\begin{figure}
  \label{cases}
  \begin{center}
  \centering
  \includegraphics[width=0.4\textwidth]{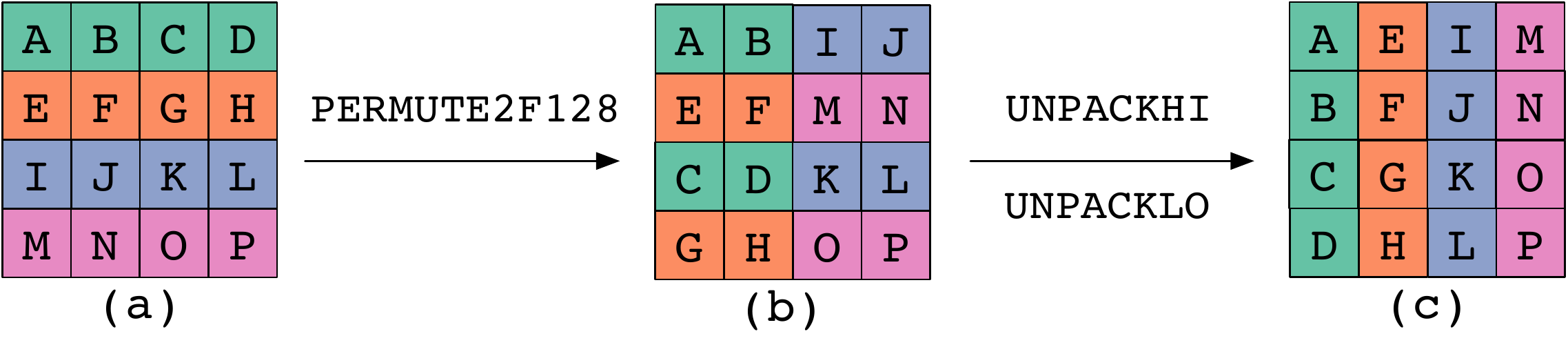}
  \caption{\label{transpose}Transpose for $double$ type using AVX-2 instructions.}
  \end{center}
\end{figure} 

Unlike previous work \cite{henretty2011data} that performs a global dimension-lifted transformation, we only need a transpose on-the-fly for each register set twice throughout the whole process.
The lower bound on the memory operations for completing a 
matrix transpose of size $vl*vl$ is $vl\log(vl)$, e.g.,
8 data reorganization instructions for $vl=4$.

In modern CPU architectures, these 8 instructions can be launched continuously in 8 cycles. However, the implementation of existing algorithm adopts lane-crossing instructions, which increases the overhead by 25\%.
Figure \ref{transpose} illustrates our improved version where the long-latency instructions are hidden by their following single-cycle instructions.
In the first stage, pairs of two vectors with distance $2$, e.g., $(A,B,C,D)$ and $(I,J,K,L)$, exchange data using the \verb=permute2f128= instruction. In the second stage, the pairs of two adjacent vectors, e.g., $(A,B,I,J)$ and $(E,F,M,N)$, swap elements by the \verb=unpackhi= or \verb=unpacklo= instruction. The total cost of the new transpose scheme is then reduced to 8 cycles.
Similarly, the transpose by using AVX-512 instructions contains three stages where the last stage consists of in-lane instructions.

\section{Evaluation}

In this section, we evaluate our proposed scheme for 1D, 2D and 3D stencils with AVX-2 and AVX-512 instructions. 

\subsection{Setup}

\begin{table}[b]
  \caption{Parameter description for stencils used in experiments}\label{parameters}
  \renewcommand\tabcolsep{10.0pt} 
  \centering\begin{tabular}{cccc}
  \hline
  Dim & Pts & Problem Size & Blocking Size\\ \hline
  1D&3    & 10240000$\times$1000 & 2000$\times$1000  \\
  1D&5    & 10240000$\times$1000 & 2000$\times$500   \\
  2D&5    & 3000$\times$3000$\times$1000 &  200$\times$200$\times$50                \\
  2D&9    & 3000$\times$3000$\times$1000 & 120$\times$128$\times$60                \\
  3D&7    &  128$\times$128$\times$128$\times$1000   &23$\times$23$\times$10                     \\
  3D&27   &   128$\times$128$\times$128$\times$1000           &  23$\times$23$\times$10                    \\ \hline
  \end{tabular}
\end{table} 
Our experiments were performed on a machine composed of two Intel Xeon Gold 6140 processors with 2.30 GHz clock speed, which owns 36 physical cores organized into two sockets. Each core contains a 32KB private L1 data cache, a  1 MB private L2 cache, and a unified 24.75MB L3 cache. AVX-512 instruction set extension is supported and it's able to conduct operations for 8 double-precision floating point data in a SIMD manner, which yields a theoretical peak performance of 73.6 GFlop/s/core (2649.6 GFlop/s in aggregate).

Since the recent tiling technique proposed by Yuan \cite{10.1145/3126908.3126920} and the nested/hybrid tiling technique (denoted as SDSL, which is the name of the software package.) presented by Henretty \cite{Henretty+:ics13} outperform the other stencil research like Pluto \cite{bondhugula2008practical,bandishti2012tiling} and Pochoir \cite{10.1145/1989493.1989508}, we take them as two counterparts of our work, which are vectorized by data multiple load and DLT methods, respectively.
All programs were compiled using the ICC compiler version 19.0.3, with the '-O3 -xHost -qopenmp -ipo' optimization flags.

The detailed parameters for stencils of various orders used in experiments are described in Table \ref{parameters}, which consists of four star stencils (1D 3-Points, 1D 5-Points, 2D 5-Points, and 3D 7-Points) and two box stencils (2D 9-Points and 3D 27-Points) corresponding to the references \cite{10.1145/3126908.3126920,Henretty+:ics13}. 
The default value of total time steps is 1000 or 200 in the references. Thus, we fix it as a larger value of 1000 in our experiments.
Other parameters of each stencil are also fine-tuned on the basis of references work to guarantee that the peak performance for all methods could be reached exactly. 
Since the performance is sensitive to the stencil parameters, significant efforts are required in automatic tuning and this will be done separately as future work.

\subsection{\label{single_thread}Sequential Block-free Results}

\begin{figure}
  \begin{center}
  \centering
  \includegraphics[width=0.5\textwidth]{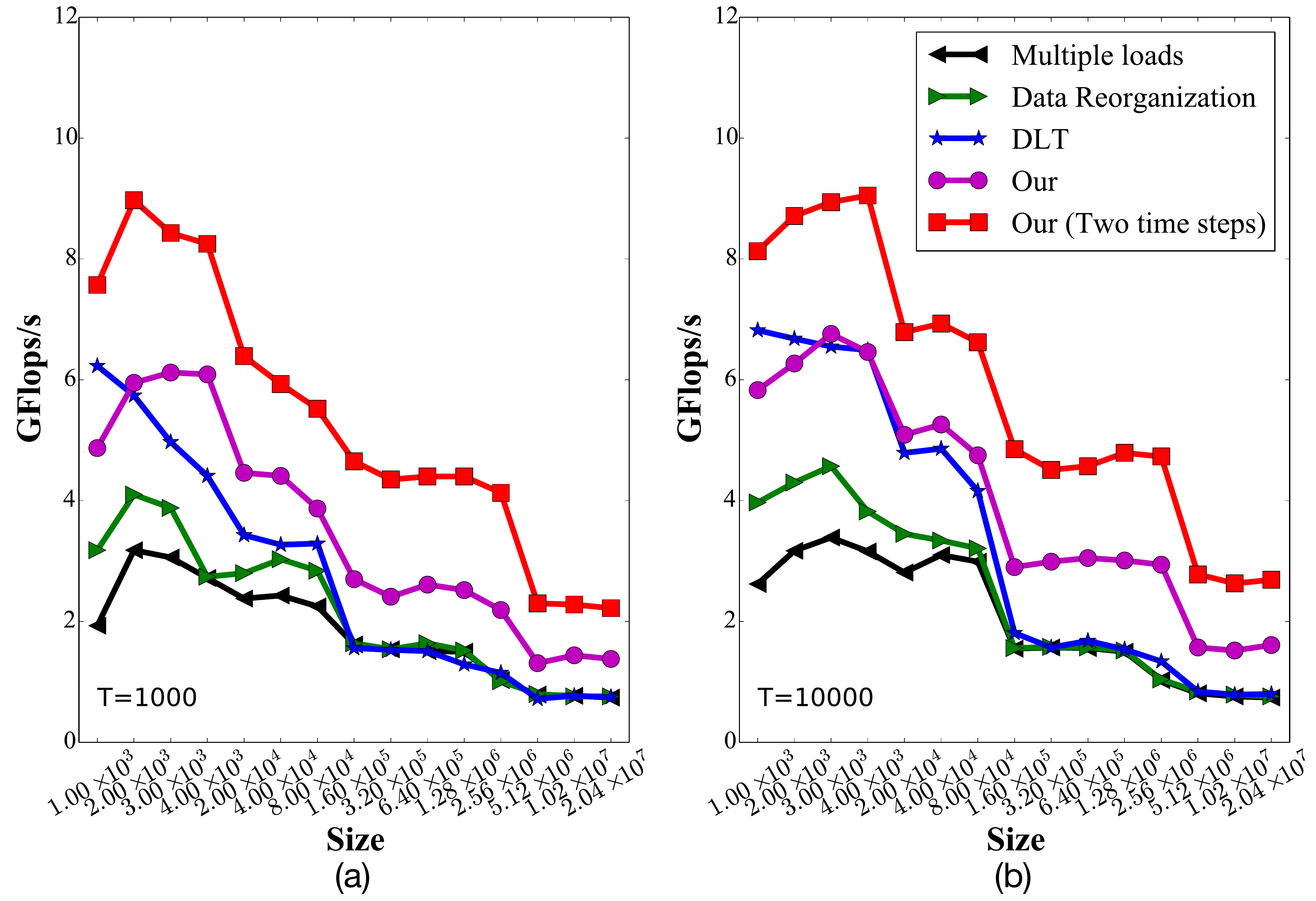}
 
  \caption{\label{bench}Absolute performance comparison for tested methods in single-thread blocking-free experiments. The results are shown separately with different total time steps.}
  \end{center}

\end{figure}

In this subsection, we present performance results of varied methods across problem sizes ranging from L1 cache to main memory with a single thread. 
The spatial and temporal blocking method is not applied to them for
examining the pure improvements on various storage levels. 
The multiple loads
and data reorganization methods represent a class of auto-vectorization in modern compilers \cite{10.1145/3126908.3126920}. DLT is the dimension-lifting transpose strategy designed by Henretty \cite{henretty2011data}.
All the methods are implemented by hand-written codes optimized with the appropriate strategies such as alignment and loop unrolling to ensure fairness.

Figure \ref{bench} shows the performance comparison of our methods with the other three methods. The results are illustrated separately in two subfigures on the basis of the total time steps $T$. It can be seen that our method updating two time steps outperforms others apparently in both experiments, which demonstrates the effectiveness of the improvement of the flop/byte ratio.
Our method without time loop unroll-and-jamming also achieves better performance results than the hand-written DLT in most cases. The performance has a decrease at the size of 1000 in L1 cache. This can be attributed to the cheaper dimension-lifting transpose operation in small size for DLT. 
The multiple loads method exhibits the worst performance among them due to the overhead caused by redundant loads. 
%

\begin{table}[htbp]
  \centering\small
  \renewcommand\tabcolsep{9.0pt} 
  \begin{threeparttable}
  \caption{\label{tab:bench}Performance improvements on different storage level in single-thread blocking-free experiments}
  \begin{tabular}{cccccc}
  \toprule Storage& {Data}  & \multirow{2}{*}{DLT} & \multirow{2}{*}{Our} &{Our }\\ 
  Level&Reorganization&&&(2 steps)\\\midrule
  \multirow{1}{*}{L1}& 1.28x  & 2.06x & 2.16x & 3.13x \\
  \multirow{1}{*}{L2}& 1.11x  & 1.37x & 1.67x& 2.07x \\
  \multirow{1}{*}{L3}& 1.01x& 0.95x& 2.02x& 2.92x \\
  \multirow{1}{*}{Memory} & 1.00x &1.01x & 1.97x & 2.96x \\\midrule
  \multirow{1}{*}{Mean}&1.11x&1.35x&1.98x&2.81x\\
  \bottomrule \end{tabular} \small
  \end{threeparttable}
\end{table}

To further investigate the effect of total time steps $T$, we perform a tenfold increase on the default value to $T=10000$, which is illustrated in Figure \ref{bench} (b). It can be observed that the performance trends of $T=10000$ are still largely consistent with the results in Figure \ref{bench} (a). However, the performance of our method falls slightly behind the DLT in L1 cache, and this performance anomaly is primarily due to the diluted dimension-lifting transpose cost by overly long time steps. Notably, only the performance of DLT in L1 cache drops gradually as problem size increases for both results in Figure \ref{bench}, which is resulted from a costly data layout transformation and indicates a potential bottleneck for cache-blocking.

\subsection{Multicore Cache-blocking Experiments}
\begin{figure}
  \begin{center}
  \centering
  \includegraphics[width=0.5\textwidth]{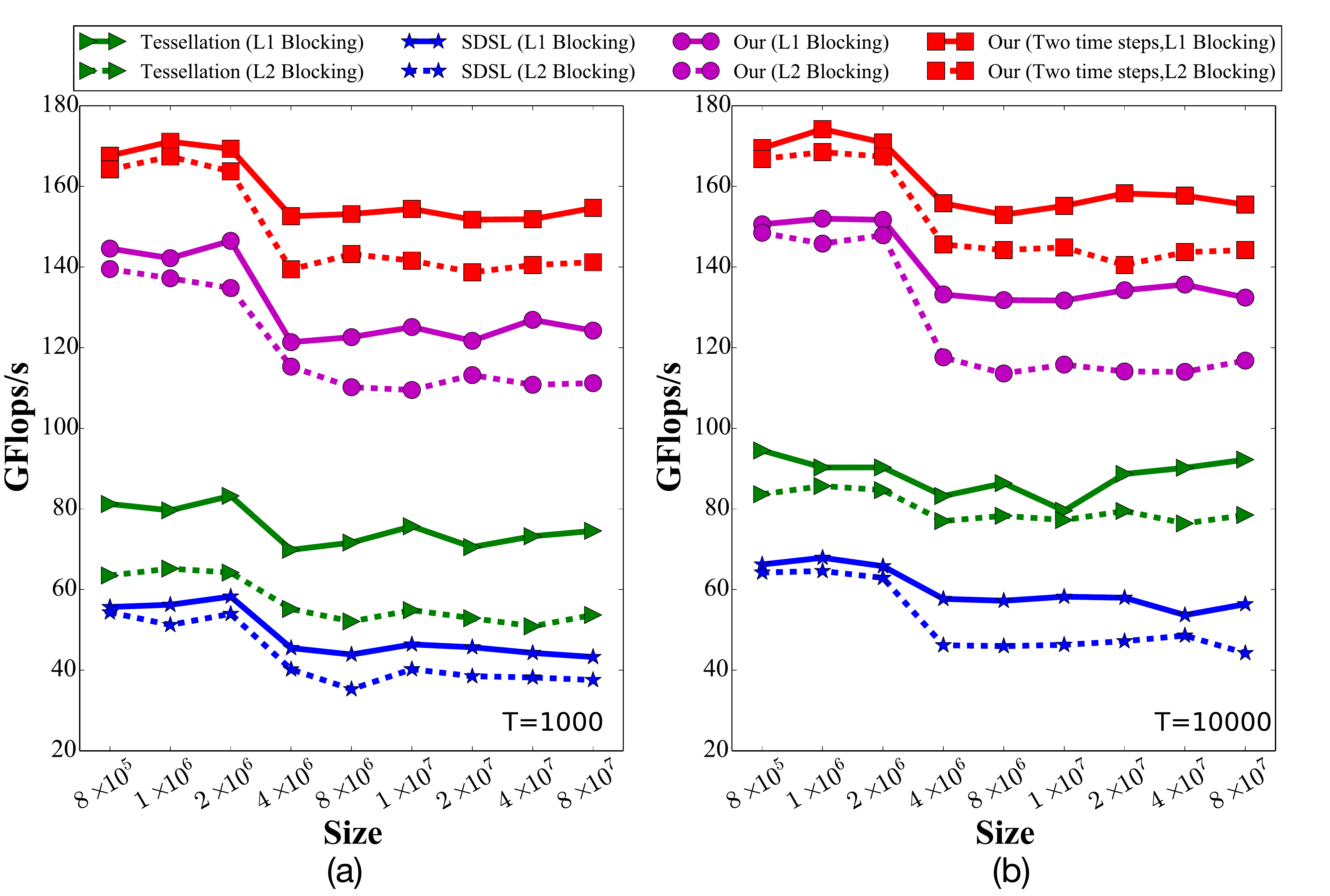}
  \caption{\label{bench_2}Absolute performance comparison for varied methods in multicore cache-blocking experiments. The results are shown separately with different total time steps.}
  \end{center}
\end{figure} 

In this subsection, we present the experiment results that exhibit the benefits of our methods with the temporal blocking and parallelization scheme. The SDSL employs a split tiling technique (nested tiling in 1D, hybrid tiling for higher dimensions) to achieve temporal blocking. The tessellate tiling technique utilized auto-vectorizing supported by the compiler  \cite{10.1145/3126908.3126920}.

The results are shown in Figure \ref{bench_2} (a) and Figure \ref{bench_2} (b) with time steps of $T=1000$ and $T=10000$ respectively. As can be seen from Figure \ref{bench_2} (a), the performance drops apparently as the problem size moves from L3 cache to the memory hierarchy, which is mainly caused by the cost of data transfers. We also further investigate the influence of the block size on performance. In the case of L1 blocking, the observed performance is higher than that with L2 blocking overall. Since the smaller stencils could be prefetched into cache directly, the performance gap between different blocks is further aggravated when the problem size lies in the memory hierarchy. Surprisingly, our method with two time steps could still take up a leading position, approximately 3.29x and 3.48x improvements are obtained compared to SDSL with L1 blocking and L2 blocking respectively. The performance of SDSL is inferior to tessellation, which is resulted from the blocking technique constrained to its data layout. Longer time steps of $T=10000$ are evaluated in Figure \ref{bench_2} (b), and similar performance trends but higher values are observed compared with Figure \ref{bench_2} (a).

Table \ref{tab:bench_2} shows the detailed performance improvements on different storage levels as before. Our method could obtain better optimization results when the problem size lies in L3 cache and memory. The speedup ranges from 2.54 to 2.76x with L1 blocking, showing that our method integrated with tiling provides a significant benefit over others on varied problem size.

\subsection{Scalability}

We also evaluated the scalabilities of our schemes and the counterparts. The detailed parameters are given in Table \ref{parameters}, where all problem sizes exceed the L3 cache. 
Since our tiling framework is the same as the tessellation scheme, the performance improvements of our method with respect to the tessellation method are fully derived from the vectorization.

\begin{figure*}
  \begin{center}
  \centering
  \includegraphics[width=1\textwidth]{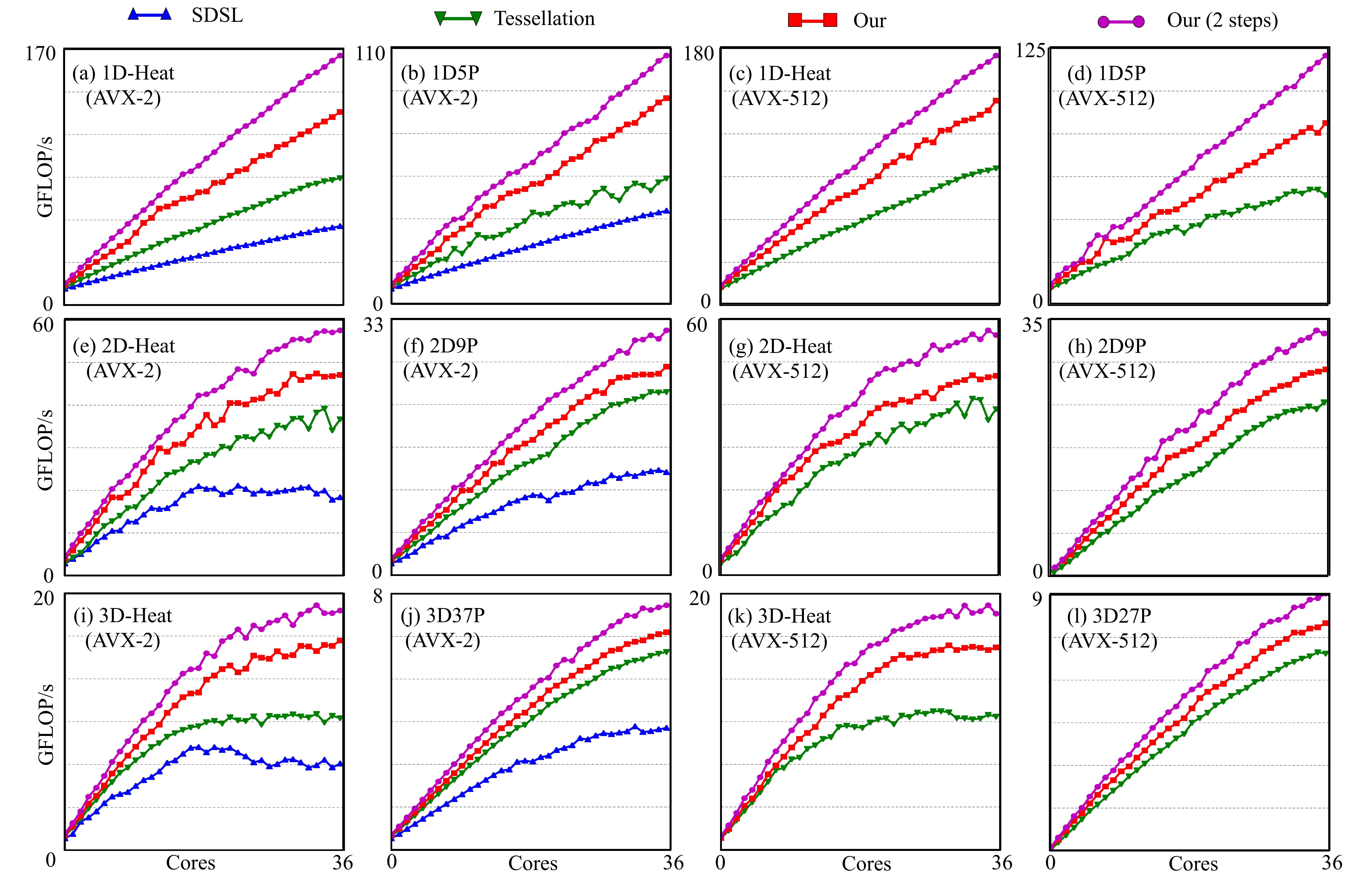}
  \caption{\label{1slope_avx512}Performance comparison for stencils of various orders with different dimensions in a multicore environment.}
  \end{center}
  \vspace{-0.3cm}
\end{figure*}

Figure \ref{1slope_avx512} illustrates the results of 1D, 2D and 3D stencils implemented with AVX-2 and AVX-512 instructions respectively. It can be observed that our method could obtain the highest performance while the SDSL performs the lowest performance. In one-dimensional stencils, all these methods achieve nearly linear scaling on both instruction sets and the proposed time loop fusion strategy provides a significant improvement. With the increase of the problem dimension, the scalability for all methods drops as a result of the inherent complexity for multidimensional stencil computations. Compared to the results implemented with AVX-2 instructions, the performance of the right half in Figure \ref{1slope_avx512} shows a slight increase.

\begin{table}[t]
  \centering\small
  \renewcommand\tabcolsep{4pt} 
  \begin{threeparttable}
  \caption{\label{tab:bench_2}Performance improvements on different storage level in multicore cache-blocking experiments}
  \begin{tabular}{lccccc}
  \toprule  &Blocking& \multirow{2}{*}{Tessellation}   & \multirow{2}{*}{Our} &{Our }\\ 
  &Level&&&(Two time steps)\\\midrule
  \multirow{2}{*}{L3 Cache} &L1& 1.43x  & 2.54x & 2.99x  \\
  &L2& 1.21x  & 2.58x & 3.01x\\
  \multirow{2}{*}{Memory}  &L1& 1.62x& 2.76x& 3.42x\\
  &L2& 1.39x &2.92x & 3.58x  \\\midrule
  \multirow{2}{*}{Mean} &L1&1.56x&2.69x&	3.29x\\
  &L2& 1.32x &2.79x & 3.48x  \\
  \bottomrule \end{tabular} \small
  \end{threeparttable}
\end{table}

\begin{table*}[t]
  \centering\small
  \renewcommand\tabcolsep{-0.2pt} 
  \begin{threeparttable}
  \caption{\label{tab:pir}Average performance improvement for different stencils}
  \begin{tabular}{cccccccccccccc}
  \toprule &\multirow{2}{*}{Method}& {1D3P}  & {2d5P} & {3D7P} &{1D3P }& {2d5P} & {3D7P} &{1D5P}  & {2d9P} & {3D27P} &{1D5P }& {2d9P} & {3D27P}\\ 
  &&(AVX-2)&(AVX-2)&(AVX-2)&(AVX-512)&(AVX-512)&(AVX-512)&(AVX-2)&(AVX-2)&(AVX-2)&(AVX-512)&(AVX-512)&(AVX-512)\\\midrule
  \multirow{1}{*}{Speedup} &SDSL& 1.00x  & 1.00x & 1.00x &-&  -  & - & 1.00x &1.00x& 1.00x  &-& - & - \\
  over&Tessellation&1.77x  &1.54x   &1.39x    & 1.00x &1.00x  & 1.00x&1.56x  &1.61x  & 1.50x&1.00x&1.00x&1.00x  \\ 
  SDSL&Our&2.77x  &2.05x & 1.85x & 1.50x & 1.24x  & 1.31x &2.37x &1.91x &1.66x&1.49x&1.26x&1.15x\\
  /Tessellation&Our$^*$&3.52x &2.26x    & 1.97x  &1.86x  & 1.34x & 1.38x &2.92x  & 2.12x & 1.76x& 1.98x & 1.39x&1.24x\\ \midrule
  \multirow{1}{*}{Speedup} &SDSL& 29.2 & 13.5 & 9.4&-&-&- &29.1 &22.0 &20.7 &-&-&-\\
  over&Tessellation& 29.8 & 24.9  &11.5    & 29.0   & 24.1 & 12.4  & 26.2 &24.1 &24.2  &24.5    & 24.0& 23.6\\ 
  {single } &Our& 30.8 &26.3  &20.1& 30.7 & 22.4  &17.7 & 31.3 &25.9&24.7&28.2&24.9&24.5\\
  core&Our$^*$& 32.1  & 26.7  & 21.2  & 31.4 & 22.3 &18.2 & 31.8 & 25.7&24.9& 28.7 & 25.6&23.8 \\
  \bottomrule \end{tabular} \small
  \begin{tablenotes}
  \item[*] Our method updated with two time steps.
  \end{tablenotes}
  \end{threeparttable}
\end{table*}

The speedups and scalabilities for high-order stencils including 1D5P, 2d9P, and 3D27P also decrease gradually from 1D to 3D. However, the overall performance falls behind the corresponding one-order results, which is resulted from complex data access patterns in high-order stencils. 
Our method could also obtain a substantial performance improvement in all experiments.

The average speedup results are given in Table \ref{tab:pir}. 
Since the stencil kernels with AVX-512 instructions are not available in the SDSL\cite{Henretty+:ics13}, the corresponding position is filled with a hyphen symbol.
Thus the comparison basis is then SDSL or Tessellation for AVX-2 or AVX-512.
Taking all stencils with AVX-2 instructions into account, remarkable performance benefits are observed from our method updating two time steps, 3.52x and 2.92x respectively for 1D3P and 1D5P. The performance improvement ranges from 1.66x to 2.77x with a mean of 2.10x, demonstrating that our vectorization scheme provides a significant benefit in a large problem size compared to the referenced work.

The speedup for each method with 36 cores is also given at the bottom of the table.
For scalability, our method obtains a 20.1x speedup while the value of DLT is only 9.4x for 3D7P, which indicates a sustainable performance for our method in multidimensional stencils. Additionally, the largest speedup in each stencil column again corresponds to the performance shown in Figure \ref{1slope_avx512}, where our method outperforms others in most cases.

\subsection{Discussion} 

In this subsection, we provide an analysis of the performance on various configurations in previous experiments to tease out the contributions from different aspects of our proposed scheme.

Sequential block-free experiments examine a variety of vectorization methods and demonstrate that our scheme with multiple time steps updating can achieve an considerable 2.81x improvement on average compared with the multiple loads method. Subsequently, the performance gains for a larger time steps are still significant and consistent with the results of the small time steps.
Moreover, the DLT method is more appropriate only on the relatively small size and long time steps, and this is partly explained by the performance penalty associated with additional dimension-lifting transpose in memory. Since the problem size ranges from L1 cache to main memory, clear insights are provided that the overall performance trends drop consistently with the various memory hierarchy.

Multicore cache-blocking experiments conduct stencil cases with 36 cores, and an average 2.69x speedup is obtained by our method on the basis of SDSL. Due to the reduced data transfers by our time loop unroll-and-jam, our method updated with two time steps achieve a further 3.29x speedup. We also study the influence of blocking size, and the results prove that appropriate L1 blocking or in-cache problem size could contribute to better performance for all methods. The overall trends are in accord with the sequential block-free experiments, and our method updated with two time steps outperforms others obviously.

The scalability experiments demonstrate that our vectorized scheme leveraging tessellate tiling successfully outperforms the referenced fastest multicore stencil work to date across a broad variety of configurations. Constrained to its specific data layout, DLT is slower than other methods. Since multidimensional or high-order stencils are more compute-intensive, more dependency data are loaded into cache while they are not fully utilized to perform their own stencil computation. Thus, the overall performance for each method falls gradually with the increasing dimensions or orders, and our method could still obtain a better performance. 

\section{Related Work}

Research on optimizing stencil computation has been intensively studied \cite{Henretty+:ics13,yuan2019tessellating,10.1145/3126908.3126920,henretty2011data,Kamil+:msp05}, and it can be broadly classified as optimization methods to improve the computation performance and enhance the data reuse.

Vectorization by using SIMD instructions is an effective way to improve computation performance for stencils. Prior work on optimizing the order of execution instructions could decrease loads/stores operations to relieve the register pressure, while only the individual element in each vector could be reused~\cite{Zumbusch:para12}.  Basu designs a vector code generation scheme to reuse several vectors in the computation process, and it is constrained to constant-coefficient and isotropic stencils \cite{7161520}. YASK \cite{yount2016yask} could improve data reuse by using common expression elimination and unrolling based on their vector-folding methods with fine-grained blocks \cite{yount2015vector}, which is less feasible for high-order complex stencils \cite{zhao2019exploiting}. Henretty proposes a new method DLT \cite{henretty2011data,Henretty+:ics13} to overcome input data alignment conflicts at the expense of a dimension-lifting transpose, which makes it infeasible to perfectly utilize the tiling technique as a result of its spatially separated data elements \cite{Krishnamoorthy+:pldi07}.
Essentially DLT can be viewed as the combination of strip-mining (1-dimensional tiling) and out-loop vectorization \cite{Henretty+:ics13}. Specifically, the original innermost loop traverses the corresponding dimension from $1$ to $N$. In DLT the loop is transformed to a depth-2 loop nest where the size of the outer loop equals the vector length $vl$ and the inner loop processes each subsequence of length $N/vl$. Note that the strip-mining was also introduced for vectorization \cite{allen2002optimizing}. However, the conventional usage is to make the size of the innermost loop be the vector length and substitute it by a vector code. In addition, the in-place matrix transpose involved in our work has also been widely studied and a kernel of 4$\times$4 matrix transpose consists of two stages basically. Hormati splits the vector register to some 128-bit lanes \cite{hormati2010macross}, and the lane-crossing instructions for $double$ incur a longer latency, typically 3 to 4 cycles. Zekri \cite{zekri2014enhancing} use in-lane instructions in four stages only for $float$ type. Springer\cite{springer2017ttc} utilize S{\scshape huffle} and P{\scshape ermute2f128} instructions for $double$ type in two stages, while it requires 8 integers as parameters.

Tiling is one of the most powerful transformation techniques
to explore the data locality of multiple loop nests~\cite{Bielecki.Palkowski:amcs16,Hartono+:ipdps10,Grosser+:ppl14}.
Notably work for stencil computations includes hyper-rectangle tiling \cite{Nguyen+:sc10,Andonov+:tpds03,Iooss+:tr15}, time skewed tiling \cite{Jin+:sc01},
diamond tiling \cite{Bandishti+:sc12},
cache oblivious Tiling  \cite{Frigo.Strumpen:ics05},
split-tiling \cite{Henretty+:ics13}
and tessellating \cite{yuan2019tessellating}.
Wonnacott and Strout present a comparison on the scalability of many existing tiling schemes \cite{Wonnacott.Strout:impact13}.
Most of these techniques are compiler transformation techniques
and this paper integrated the new proposed layout with the tessellation scheme for simplifying the implementation.
For stencil computations, a variety of auto-tuning frameworks \cite{christen2011patus,gysi2015modesto,Rocha+:ccpe17} have been presented by using varied hyper-rectangular tiles to exploit data reuse alone. However, redundant computations are involved in these work to resolve the introduced inter-tile dependencies that hinder the concurrent execution of shaped tiles on different cores.  

\section{Conclusion}

In this paper, we propose a novel transpose layout to overcome the input data alignment conflicts efficiently for vectorization. A time loop unroll-and-jam strategy with in-register multiple time steps processing is designed on the basis of the proposed transpose layout.
Furthermore, we describe how the proposed vectorization scheme is integrated with a tessellate tiling framework for enhancing data reuse and concurrency. With the qualitative analysis and quantitative experiments, we demonstrate that significant performance improvements are achieved by our vectorization scheme over state-of-the-art products such as Intel’s ICC and recent work \cite{Henretty+:ics13,10.1145/3126908.3126920}. Experimental results provide evidence that our vectorization scheme incorporated with tessellate tiling could obtain a linear scaling character and reach a 3.29x improvement over SDSL for one-dimensional stencils.

\section{Acknowledgements} 
This work is supported by the National Key R\&D Program of China under Grant No.2016YFB0200803. 


\bibliographystyle{ACM-Reference-Format}
\bibliography{mybibfile}


\begin{thebibliography}{40}


\ifx \showCODEN    \undefined \def \showCODEN     #1{\unskip}     \fi
\ifx \showDOI      \undefined \def \showDOI       #1{#1}\fi
\ifx \showISBNx    \undefined \def \showISBNx     #1{\unskip}     \fi
\ifx \showISBNxiii \undefined \def \showISBNxiii  #1{\unskip}     \fi
\ifx \showISSN     \undefined \def \showISSN      #1{\unskip}     \fi
\ifx \showLCCN     \undefined \def \showLCCN      #1{\unskip}     \fi
\ifx \shownote     \undefined \def \shownote      #1{#1}          \fi
\ifx \showarticletitle \undefined \def \showarticletitle #1{#1}   \fi
\ifx \showURL      \undefined \def \showURL       {\relax}        \fi
\providecommand\bibfield[2]{#2}
\providecommand\bibinfo[2]{#2}
\providecommand\natexlab[1]{#1}
\providecommand\showeprint[2][]{arXiv:#2}

\bibitem[\protect\citeauthoryear{Allen and Kennedy}{Allen and Kennedy}{2002}]%
        {allen2002optimizing}
\bibfield{author}{\bibinfo{person}{Randy Allen} {and} \bibinfo{person}{Ken
  Kennedy}.} \bibinfo{year}{2002}\natexlab{}.
\newblock \bibinfo{booktitle}{\emph{Optimizing compilers for modern
  architectures: a dependence-based approach}}.
\newblock \bibinfo{publisher}{Taylor \& Francis US}.
\newblock


\bibitem[\protect\citeauthoryear{Andonov, Balev, Rajopadhye, and Yanev}{Andonov
  et~al\mbox{.}}{2003}]%
        {Andonov+:tpds03}
\bibfield{author}{\bibinfo{person}{R. Andonov}, \bibinfo{person}{S. Balev},
  \bibinfo{person}{S. Rajopadhye}, {and} \bibinfo{person}{N. Yanev}.}
  \bibinfo{year}{2003}\natexlab{}.
\newblock \showarticletitle{Optimal semi-oblique tiling}.
\newblock \bibinfo{journal}{\emph{IEEE Transactions on Parallel and Distributed
  Systems}} \bibinfo{volume}{14}, \bibinfo{number}{9} (\bibinfo{year}{2003}),
  \bibinfo{pages}{944--960}.
\newblock
\showISSN{1045-9219}


\bibitem[\protect\citeauthoryear{Asanovic, Bodik, Demmel,
  et~al\mbox{.}}{Asanovic et~al\mbox{.}}{2008}]%
        {asanovic2008parallel}
\bibfield{author}{\bibinfo{person}{Krste Asanovic}, \bibinfo{person}{Ras
  Bodik}, \bibinfo{person}{James Demmel}, {et~al\mbox{.}}}
  \bibinfo{year}{2008}\natexlab{}.
\newblock \showarticletitle{The parallel computing laboratory at UC Berkeley: A
  research agenda based on the Berkeley view}.
\newblock \bibinfo{journal}{\emph{University of California, Berkeley, Tech.
  Rep}} (\bibinfo{year}{2008}).
\newblock


\bibitem[\protect\citeauthoryear{Bandishti, Pananilath, and
  Bondhugula}{Bandishti et~al\mbox{.}}{2012a}]%
        {bandishti2012tiling}
\bibfield{author}{\bibinfo{person}{Vinayaka Bandishti}, \bibinfo{person}{Irshad
  Pananilath}, {and} \bibinfo{person}{Uday Bondhugula}.}
  \bibinfo{year}{2012}\natexlab{a}.
\newblock \showarticletitle{Tiling stencil computations to maximize
  parallelism}. In \bibinfo{booktitle}{\emph{SC'12}}. IEEE,
  \bibinfo{pages}{1--11}.
\newblock


\bibitem[\protect\citeauthoryear{Bandishti, Pananilath, and
  Bondhugula}{Bandishti et~al\mbox{.}}{2012b}]%
        {Bandishti+:sc12}
\bibfield{author}{\bibinfo{person}{V. Bandishti}, \bibinfo{person}{I.
  Pananilath}, {and} \bibinfo{person}{U. Bondhugula}.}
  \bibinfo{year}{2012}\natexlab{b}.
\newblock \showarticletitle{Tiling stencil computations to maximize
  parallelism} \emph{(\bibinfo{series}{SC '12})}. \bibinfo{pages}{1--11}.
\newblock
\showISSN{2167-4329}


\bibitem[\protect\citeauthoryear{{Basu}, {Hall}, {Williams}, {Straalen},
  {Oliker}, and {Colella}}{{Basu} et~al\mbox{.}}{2015}]%
        {7161520}
\bibfield{author}{\bibinfo{person}{P. {Basu}}, \bibinfo{person}{M. {Hall}},
  \bibinfo{person}{S. {Williams}}, \bibinfo{person}{B.~V. {Straalen}},
  \bibinfo{person}{L. {Oliker}}, {and} \bibinfo{person}{P. {Colella}}.}
  \bibinfo{year}{2015}\natexlab{}.
\newblock \showarticletitle{Compiler-Directed Transformation for Higher-Order
  Stencils}. In \bibinfo{booktitle}{\emph{2015 IEEE International Parallel and
  Distributed Processing Symposium}}. \bibinfo{pages}{313--323}.
\newblock


\bibitem[\protect\citeauthoryear{Bielecki and Palkowski}{Bielecki and
  Palkowski}{2016}]%
        {Bielecki.Palkowski:amcs16}
\bibfield{author}{\bibinfo{person}{Wlodzimierz Bielecki} {and}
  \bibinfo{person}{Marek Palkowski}.} \bibinfo{year}{2016}\natexlab{}.
\newblock \showarticletitle{Tiling arbitrarily nested loops by means of the
  transitive closure of dependence graphs}.
\newblock \bibinfo{journal}{\emph{AMCS : International Journal of Applied
  Mathematics and Computer Science}} \bibinfo{volume}{26}, \bibinfo{number}{4}
  (\bibinfo{year}{2016}), \bibinfo{pages}{919--939}.
\newblock
\urldef\tempurl%
\url{https://doi.org/10.1515/amcs-2016-0065}
\showDOI{\tempurl}


\bibitem[\protect\citeauthoryear{Bondhugula, Hartono, Ramanujam, and
  Sadayappan}{Bondhugula et~al\mbox{.}}{2008}]%
        {bondhugula2008practical}
\bibfield{author}{\bibinfo{person}{Uday Bondhugula}, \bibinfo{person}{Albert
  Hartono}, \bibinfo{person}{Jagannathan Ramanujam}, {and}
  \bibinfo{person}{Ponnuswamy Sadayappan}.} \bibinfo{year}{2008}\natexlab{}.
\newblock \showarticletitle{A practical automatic polyhedral parallelizer and
  locality optimizer}. In \bibinfo{booktitle}{\emph{PLDI 08}}.
\newblock


\bibitem[\protect\citeauthoryear{Christen, Schenk, and Burkhart}{Christen
  et~al\mbox{.}}{[n. d.]}]%
        {christen2011patus}
\bibfield{author}{\bibinfo{person}{Matthias Christen}, \bibinfo{person}{Olaf
  Schenk}, {and} \bibinfo{person}{Helmar Burkhart}.} \bibinfo{year}{[n.
  d.]}\natexlab{}.
\newblock \showarticletitle{Patus: A code generation and autotuning framework
  for parallel iterative stencil computations on modern microarchitectures}. In
  \bibinfo{booktitle}{\emph{IPDPS 2011}}. IEEE.
\newblock


\bibitem[\protect\citeauthoryear{Datta, Kamil, Williams, Oliker, Shalf, and
  Yelick}{Datta et~al\mbox{.}}{2009}]%
        {datta2009optimization}
\bibfield{author}{\bibinfo{person}{Kaushik Datta}, \bibinfo{person}{Shoaib
  Kamil}, \bibinfo{person}{Samuel Williams}, \bibinfo{person}{Leonid Oliker},
  \bibinfo{person}{John Shalf}, {and} \bibinfo{person}{Katherine Yelick}.}
  \bibinfo{year}{2009}\natexlab{}.
\newblock \showarticletitle{Optimization and performance modeling of stencil
  computations on modern microprocessors}.
\newblock \bibinfo{journal}{\emph{SIAM review}} \bibinfo{volume}{51},
  \bibinfo{number}{1} (\bibinfo{year}{2009}), \bibinfo{pages}{129--159}.
\newblock


\bibitem[\protect\citeauthoryear{Dursun, Kunaseth, Nomura, Chame, Lucas, Chen,
  Hall, Kalia, Nakano, and Vashishta}{Dursun et~al\mbox{.}}{2012}]%
        {Dursun+:jsupercomputing12}
\bibfield{author}{\bibinfo{person}{Hikmet Dursun}, \bibinfo{person}{Manaschai
  Kunaseth}, \bibinfo{person}{Ken-ichi Nomura}, \bibinfo{person}{Jacqueline
  Chame}, \bibinfo{person}{RobertF. Lucas}, \bibinfo{person}{Chun Chen},
  \bibinfo{person}{Mary Hall}, \bibinfo{person}{RajivK. Kalia},
  \bibinfo{person}{Aiichiro Nakano}, {and} \bibinfo{person}{Priya Vashishta}.}
  \bibinfo{year}{2012}\natexlab{}.
\newblock \showarticletitle{Hierarchical parallelization and optimization of
  high-order stencil computations on multicore clusters}.
\newblock \bibinfo{journal}{\emph{The Journal of Supercomputing}}
  \bibinfo{volume}{62}, \bibinfo{number}{2} (\bibinfo{year}{2012}),
  \bibinfo{pages}{946--966}.
\newblock
\showISSN{0920-8542}
\urldef\tempurl%
\url{https://doi.org/10.1007/s11227-012-0764-z}
\showDOI{\tempurl}


\bibitem[\protect\citeauthoryear{D\"{u}tsch, Djelassi, Haidl, and
  Gorlatch}{D\"{u}tsch et~al\mbox{.}}{2014}]%
        {Dutsch+:wosc14}
\bibfield{author}{\bibinfo{person}{Fabian D\"{u}tsch}, \bibinfo{person}{Karim
  Djelassi}, \bibinfo{person}{Michael Haidl}, {and} \bibinfo{person}{Sergei
  Gorlatch}.} \bibinfo{year}{2014}\natexlab{}.
\newblock \showarticletitle{HLSF: A High-Level; C++-Based Framework for Stencil
  Computations on Accelerators} \emph{(\bibinfo{series}{WOSC '14})}.
  \bibinfo{pages}{41--4}.
\newblock
\showISBNx{978-1-4503-2308-6}


\bibitem[\protect\citeauthoryear{Falke, Merz, and Sinz}{Falke
  et~al\mbox{.}}{2013}]%
        {falke2013extending}
\bibfield{author}{\bibinfo{person}{Stephan Falke}, \bibinfo{person}{Florian
  Merz}, {and} \bibinfo{person}{Carsten Sinz}.}
  \bibinfo{year}{2013}\natexlab{}.
\newblock \showarticletitle{Extending the Theory of Arrays: memset, memcpy, and
  Beyond}. In \bibinfo{booktitle}{\emph{Working Conference on Verified
  Software: Theories, Tools, and Experiments}}. Springer,
  \bibinfo{pages}{108--128}.
\newblock


\bibitem[\protect\citeauthoryear{Frigo and Strumpen}{Frigo and
  Strumpen}{2005}]%
        {Frigo.Strumpen:ics05}
\bibfield{author}{\bibinfo{person}{Matteo Frigo} {and} \bibinfo{person}{Volker
  Strumpen}.} \bibinfo{year}{2005}\natexlab{}.
\newblock \showarticletitle{Cache oblivious stencil computations}
  \emph{(\bibinfo{series}{ICS '05})}. \bibinfo{pages}{361--366}.
\newblock


\bibitem[\protect\citeauthoryear{Grosser, Verdoolaege, Cohen, and
  Sadayappan}{Grosser et~al\mbox{.}}{2014}]%
        {Grosser+:ppl14}
\bibfield{author}{\bibinfo{person}{Tobias Grosser}, \bibinfo{person}{Sven
  Verdoolaege}, \bibinfo{person}{Albert Cohen}, {and} \bibinfo{person}{P.
  Sadayappan}.} \bibinfo{year}{2014}\natexlab{}.
\newblock \showarticletitle{The Relation Between Diamond Tiling and Hexagonal
  Tiling}.
\newblock \bibinfo{journal}{\emph{Parallel Processing Letters}}
  \bibinfo{volume}{24}, \bibinfo{number}{03} (\bibinfo{year}{2014}).
\newblock


\bibitem[\protect\citeauthoryear{Gysi, Grosser, and Hoefler}{Gysi
  et~al\mbox{.}}{2015}]%
        {gysi2015modesto}
\bibfield{author}{\bibinfo{person}{Tobias Gysi}, \bibinfo{person}{Tobias
  Grosser}, {and} \bibinfo{person}{Torsten Hoefler}.}
  \bibinfo{year}{2015}\natexlab{}.
\newblock \showarticletitle{Modesto: Data-centric analytic optimization of
  complex stencil programs on heterogeneous architectures}. In
  \bibinfo{booktitle}{\emph{ICS 2015}}. \bibinfo{pages}{177--186}.
\newblock


\bibitem[\protect\citeauthoryear{Hartono, Baskaran, Ramanujam, and
  Sadayappan}{Hartono et~al\mbox{.}}{2010}]%
        {Hartono+:ipdps10}
\bibfield{author}{\bibinfo{person}{A. Hartono}, \bibinfo{person}{M.~M.
  Baskaran}, \bibinfo{person}{J. Ramanujam}, {and} \bibinfo{person}{P.
  Sadayappan}.} \bibinfo{year}{2010}\natexlab{}.
\newblock \showarticletitle{DynTile: Parametric tiled loop generation for
  parallel execution on multicore processors} \emph{(\bibinfo{series}{IPDPS
  '10})}. \bibinfo{pages}{1--12}.
\newblock


\bibitem[\protect\citeauthoryear{Henretty, Stock, Pouchet, Franchetti,
  Ramanujam, and Sadayappan}{Henretty et~al\mbox{.}}{2011}]%
        {henretty2011data}
\bibfield{author}{\bibinfo{person}{Tom Henretty}, \bibinfo{person}{Kevin
  Stock}, \bibinfo{person}{Louis-No{\"e}l Pouchet}, \bibinfo{person}{Franz
  Franchetti}, \bibinfo{person}{J Ramanujam}, {and} \bibinfo{person}{P
  Sadayappan}.} \bibinfo{year}{2011}\natexlab{}.
\newblock \showarticletitle{Data layout transformation for stencil computations
  on short-vector simd architectures}. In
  \bibinfo{booktitle}{\emph{International Conference on Compiler
  Construction}}. Springer, \bibinfo{pages}{225--245}.
\newblock


\bibitem[\protect\citeauthoryear{Henretty, Veras, Franchetti, Pouchet,
  Ramanujam, and Sadayappan}{Henretty et~al\mbox{.}}{2013}]%
        {Henretty+:ics13}
\bibfield{author}{\bibinfo{person}{Tom Henretty}, \bibinfo{person}{Richard
  Veras}, \bibinfo{person}{Franz Franchetti}, \bibinfo{person}{Louis-No\"{e}l
  Pouchet}, \bibinfo{person}{J. Ramanujam}, {and} \bibinfo{person}{P.
  Sadayappan}.} \bibinfo{year}{2013}\natexlab{}.
\newblock \showarticletitle{A Stencil Compiler for Short-vector SIMD
  Architectures} \emph{(\bibinfo{series}{ICS '13})}. \bibinfo{pages}{13--24}.
\newblock
\showISBNx{978-1-4503-2130-3}


\bibitem[\protect\citeauthoryear{Hormati, Choi, Woh, Kudlur, Rabbah, Mudge, and
  Mahlke}{Hormati et~al\mbox{.}}{2010}]%
        {hormati2010macross}
\bibfield{author}{\bibinfo{person}{Amir~H Hormati}, \bibinfo{person}{Yoonseo
  Choi}, \bibinfo{person}{Mark Woh}, \bibinfo{person}{Manjunath Kudlur},
  \bibinfo{person}{Rodric Rabbah}, \bibinfo{person}{Trevor Mudge}, {and}
  \bibinfo{person}{Scott Mahlke}.} \bibinfo{year}{2010}\natexlab{}.
\newblock \showarticletitle{MacroSS: macro-SIMDization of streaming
  applications}.
\newblock \bibinfo{journal}{\emph{ACM SIGARCH computer architecture news}}
  \bibinfo{volume}{38}, \bibinfo{number}{1} (\bibinfo{year}{2010}),
  \bibinfo{pages}{285--296}.
\newblock


\bibitem[\protect\citeauthoryear{Iooss, Rajopadhye, Alias, and Zou}{Iooss
  et~al\mbox{.}}{2015}]%
        {Iooss+:tr15}
\bibfield{author}{\bibinfo{person}{Guillaume Iooss}, \bibinfo{person}{Sanjay
  Rajopadhye}, \bibinfo{person}{Christophe Alias}, {and} \bibinfo{person}{Yun
  Zou}.} \bibinfo{year}{2015}\natexlab{}.
\newblock \bibinfo{booktitle}{\emph{Mono-parametric Tiling is a Polyhedral
  Transformation}}.
\newblock \bibinfo{type}{Research Report}.
\newblock


\bibitem[\protect\citeauthoryear{Jin, Mellor-Crummey, and Fowler}{Jin
  et~al\mbox{.}}{2001}]%
        {Jin+:sc01}
\bibfield{author}{\bibinfo{person}{Guohua Jin}, \bibinfo{person}{John
  Mellor-Crummey}, {and} \bibinfo{person}{Robert Fowler}.}
  \bibinfo{year}{2001}\natexlab{}.
\newblock \showarticletitle{Increasing Temporal Locality with Skewing and
  Recursive Blocking} \emph{(\bibinfo{series}{SC '01})}.
  \bibinfo{pages}{43--43}.
\newblock
\showISBNx{1-58113-293-X}


\bibitem[\protect\citeauthoryear{Kamil, Husbands, Oliker, Shalf, and
  Yelick}{Kamil et~al\mbox{.}}{2005}]%
        {Kamil+:msp05}
\bibfield{author}{\bibinfo{person}{Shoaib Kamil}, \bibinfo{person}{Parry
  Husbands}, \bibinfo{person}{Leonid Oliker}, \bibinfo{person}{John Shalf},
  {and} \bibinfo{person}{Katherine Yelick}.} \bibinfo{year}{2005}\natexlab{}.
\newblock \showarticletitle{Impact of Modern Memory Subsystems on Cache
  Optimizations for Stencil Computations} \emph{(\bibinfo{series}{MSP '05})}.
  \bibinfo{pages}{36--43}.
\newblock
\showISBNx{1-59593-147-3}


\bibitem[\protect\citeauthoryear{Krishnamoorthy, Baskaran, Bondhugula,
  Ramanujam, Rountev, and Sadayappan}{Krishnamoorthy et~al\mbox{.}}{2007}]%
        {Krishnamoorthy+:pldi07}
\bibfield{author}{\bibinfo{person}{Sriram Krishnamoorthy},
  \bibinfo{person}{Muthu Baskaran}, \bibinfo{person}{Uday Bondhugula},
  \bibinfo{person}{J. Ramanujam}, \bibinfo{person}{Atanas Rountev}, {and}
  \bibinfo{person}{P Sadayappan}.} \bibinfo{year}{2007}\natexlab{}.
\newblock \showarticletitle{Effective Automatic Parallelization of Stencil
  Computations} \emph{(\bibinfo{series}{PLDI '07})}. \bibinfo{pages}{235--244}.
\newblock
\showISBNx{978-1-59593-633-2}


\bibitem[\protect\citeauthoryear{Li, Shang, Zhang, et~al\mbox{.}}{Li
  et~al\mbox{.}}{2019}]%
        {li2019openkmc}
\bibfield{author}{\bibinfo{person}{Kun Li}, \bibinfo{person}{Honghui Shang},
  \bibinfo{person}{Yunquan Zhang}, {et~al\mbox{.}}}
  \bibinfo{year}{2019}\natexlab{}.
\newblock \showarticletitle{OpenKMC: a KMC design for hundred-billion-atom
  simulation using millions of cores on Sunway Taihulight}. In
  \bibinfo{booktitle}{\emph{SC'19}}.
\newblock


\bibitem[\protect\citeauthoryear{Luo, Tan, Mo, and Sun}{Luo
  et~al\mbox{.}}{2015}]%
        {Luo+:ics15}
\bibfield{author}{\bibinfo{person}{Yulong Luo}, \bibinfo{person}{Guangming
  Tan}, \bibinfo{person}{Zeyao Mo}, {and} \bibinfo{person}{Ninghui Sun}.}
  \bibinfo{year}{2015}\natexlab{}.
\newblock \showarticletitle{FAST: A Fast Stencil Autotuning Framework Based On
  An Optimal-solution Space Model} \emph{(\bibinfo{series}{ICS '15})}.
  \bibinfo{pages}{187--196}.
\newblock
\showISBNx{978-1-4503-3559-1}


\bibitem[\protect\citeauthoryear{Nguyen, Satish, Chhugani, Kim, and
  Dubey}{Nguyen et~al\mbox{.}}{2010}]%
        {Nguyen+:sc10}
\bibfield{author}{\bibinfo{person}{A. Nguyen}, \bibinfo{person}{N. Satish},
  \bibinfo{person}{J. Chhugani}, \bibinfo{person}{C. Kim}, {and}
  \bibinfo{person}{P. Dubey}.} \bibinfo{year}{2010}\natexlab{}.
\newblock \showarticletitle{3.5-D Blocking Optimization for Stencil
  Computations on Modern CPUs and GPUs} \emph{(\bibinfo{series}{SC '10})}.
  \bibinfo{pages}{1--13}.
\newblock
\showISSN{2167-4329}


\bibitem[\protect\citeauthoryear{Rawat, Rastello, Sukumaran-Rajam, Pouchet,
  Rountev, and Sadayappan}{Rawat et~al\mbox{.}}{[n. d.]}]%
        {rawat2018register}
\bibfield{author}{\bibinfo{person}{Prashant~Singh Rawat},
  \bibinfo{person}{Fabrice Rastello}, \bibinfo{person}{Aravind
  Sukumaran-Rajam}, \bibinfo{person}{Louis-No{\"e}l Pouchet},
  \bibinfo{person}{Atanas Rountev}, {and} \bibinfo{person}{P Sadayappan}.}
  \bibinfo{year}{[n. d.]}\natexlab{}.
\newblock \showarticletitle{Register optimizations for stencils on GPUs}. In
  \bibinfo{booktitle}{\emph{PPOPP'2018}}. \bibinfo{pages}{168--182}.
\newblock


\bibitem[\protect\citeauthoryear{{Rawat}, {Sukumaran-Rajam}, {Rountev},
  {Rastello}, {Pouchet}, and {Sadayappan}}{{Rawat} et~al\mbox{.}}{2018}]%
        {8665800}
\bibfield{author}{\bibinfo{person}{P.~S. {Rawat}}, \bibinfo{person}{A.
  {Sukumaran-Rajam}}, \bibinfo{person}{A. {Rountev}}, \bibinfo{person}{F.
  {Rastello}}, \bibinfo{person}{L. {Pouchet}}, {and} \bibinfo{person}{P.
  {Sadayappan}}.} \bibinfo{year}{2018}\natexlab{}.
\newblock \showarticletitle{Associative Instruction Reordering to Alleviate
  Register Pressure}. In \bibinfo{booktitle}{\emph{SC'18}}.
  \bibinfo{pages}{590--602}.
\newblock


\bibitem[\protect\citeauthoryear{Rocha, Pereira, Ramos, and Góes}{Rocha
  et~al\mbox{.}}{2017}]%
        {Rocha+:ccpe17}
\bibfield{author}{\bibinfo{person}{Rodrigo C.~O. Rocha},
  \bibinfo{person}{Alyson~D. Pereira}, \bibinfo{person}{Luiz Ramos}, {and}
  \bibinfo{person}{Luís F.~W. Góes}.} \bibinfo{year}{2017}\natexlab{}.
\newblock \showarticletitle{TOAST: Automatic tiling for iterative stencil
  computations on GPUs}.
\newblock \bibinfo{journal}{\emph{Concurrency and Computation: Practice and
  Experience}} (\bibinfo{year}{2017}).
\newblock
\showISSN{1532-0634}


\bibitem[\protect\citeauthoryear{Springer, Hammond, and Bientinesi}{Springer
  et~al\mbox{.}}{2017}]%
        {springer2017ttc}
\bibfield{author}{\bibinfo{person}{Paul Springer}, \bibinfo{person}{Jeff~R
  Hammond}, {and} \bibinfo{person}{Paolo Bientinesi}.}
  \bibinfo{year}{2017}\natexlab{}.
\newblock \showarticletitle{TTC: A high-performance compiler for tensor
  transpositions}.
\newblock \bibinfo{journal}{\emph{ACM Transactions on Mathematical Software
  (TOMS)}} \bibinfo{volume}{44}, \bibinfo{number}{2} (\bibinfo{year}{2017}),
  \bibinfo{pages}{1--21}.
\newblock


\bibitem[\protect\citeauthoryear{Tang, Chowdhury, Kuszmaul, Luk, and
  Leiserson}{Tang et~al\mbox{.}}{2011}]%
        {10.1145/1989493.1989508}
\bibfield{author}{\bibinfo{person}{Yuan Tang}, \bibinfo{person}{Rezaul~Alam
  Chowdhury}, \bibinfo{person}{Bradley~C. Kuszmaul}, \bibinfo{person}{Chi-Keung
  Luk}, {and} \bibinfo{person}{Charles~E. Leiserson}.}
  \bibinfo{year}{2011}\natexlab{}.
\newblock \showarticletitle{The Pochoir Stencil Compiler}. In
  \bibinfo{booktitle}{\emph{SPAA'11}}. \bibinfo{publisher}{ACM},
  \bibinfo{address}{New York, NY, USA}, 12.
\newblock
\showISBNx{9781450307437}
\urldef\tempurl%
\url{https://doi.org/10.1145/1989493.1989508}
\showDOI{\tempurl}


\bibitem[\protect\citeauthoryear{Wonnacott and Strout}{Wonnacott and
  Strout}{2013}]%
        {Wonnacott.Strout:impact13}
\bibfield{author}{\bibinfo{person}{David~G Wonnacott} {and}
  \bibinfo{person}{Michelle~Mills Strout}.} \bibinfo{year}{2013}\natexlab{}.
\newblock \showarticletitle{On the scalability of loop tiling techniques}.
\newblock \bibinfo{journal}{\emph{IMPACT 2013}} (\bibinfo{year}{2013}).
\newblock


\bibitem[\protect\citeauthoryear{Yount}{Yount}{2015}]%
        {yount2015vector}
\bibfield{author}{\bibinfo{person}{Charles Yount}.}
  \bibinfo{year}{2015}\natexlab{}.
\newblock \showarticletitle{Vector Folding: improving stencil performance via
  multi-dimensional SIMD-vector representation}. In
  \bibinfo{booktitle}{\emph{2015 IEEE 17th International Conference on High
  Performance Computing and Communications, 2015 IEEE 7th International
  Symposium on Cyberspace Safety and Security, and 2015 IEEE 12th International
  Conference on Embedded Software and Systems}}. IEEE,
  \bibinfo{pages}{865--870}.
\newblock


\bibitem[\protect\citeauthoryear{Yount, Tobin, Breuer, and Duran}{Yount
  et~al\mbox{.}}{2016}]%
        {yount2016yask}
\bibfield{author}{\bibinfo{person}{Charles Yount}, \bibinfo{person}{Josh
  Tobin}, \bibinfo{person}{Alexander Breuer}, {and} \bibinfo{person}{Alejandro
  Duran}.} \bibinfo{year}{2016}\natexlab{}.
\newblock \showarticletitle{YASK—Yet another stencil kernel: A framework for
  HPC stencil code-generation and tuning}. In \bibinfo{booktitle}{\emph{WOLFHPC
  2016}}. IEEE, \bibinfo{pages}{30--39}.
\newblock


\bibitem[\protect\citeauthoryear{Yuan, Huang, Zhang, and Cao}{Yuan
  et~al\mbox{.}}{2019}]%
        {yuan2019tessellating}
\bibfield{author}{\bibinfo{person}{Liang Yuan}, \bibinfo{person}{Shan Huang},
  \bibinfo{person}{Yunquan Zhang}, {and} \bibinfo{person}{Hang Cao}.}
  \bibinfo{year}{2019}\natexlab{}.
\newblock \showarticletitle{Tessellating Star Stencils}. In
  \bibinfo{booktitle}{\emph{Proceedings of the 48th International Conference on
  Parallel Processing}}. \bibinfo{pages}{1--10}.
\newblock


\bibitem[\protect\citeauthoryear{Yuan, Zhang, Guo, and Huang}{Yuan
  et~al\mbox{.}}{2017}]%
        {10.1145/3126908.3126920}
\bibfield{author}{\bibinfo{person}{Liang Yuan}, \bibinfo{person}{Yunquan
  Zhang}, \bibinfo{person}{Peng Guo}, {and} \bibinfo{person}{Shan Huang}.}
  \bibinfo{year}{2017}\natexlab{}.
\newblock \showarticletitle{Tessellating Stencils}. In
  \bibinfo{booktitle}{\emph{SC'17}}. \bibinfo{publisher}{ACM},
  \bibinfo{address}{New York, NY, USA}, Article \bibinfo{articleno}{Article
  49}, \bibinfo{numpages}{13}~pages.
\newblock
\showISBNx{9781450351140}
\urldef\tempurl%
\url{https://doi.org/10.1145/3126908.3126920}
\showDOI{\tempurl}


\bibitem[\protect\citeauthoryear{Zekri}{Zekri}{2014}]%
        {zekri2014enhancing}
\bibfield{author}{\bibinfo{person}{Ahmed~Sherif Zekri}.}
  \bibinfo{year}{2014}\natexlab{}.
\newblock \showarticletitle{Enhancing the matrix transpose operation using
  intel avx instruction set extension}.
\newblock \bibinfo{journal}{\emph{International Journal of Computer Science \&
  Information Technology}} \bibinfo{volume}{6}, \bibinfo{number}{3}
  (\bibinfo{year}{2014}), \bibinfo{pages}{67}.
\newblock


\bibitem[\protect\citeauthoryear{Zhao, Basu, Williams, Hall, and Johansen}{Zhao
  et~al\mbox{.}}{2019}]%
        {zhao2019exploiting}
\bibfield{author}{\bibinfo{person}{Tuowen Zhao}, \bibinfo{person}{Protonu
  Basu}, \bibinfo{person}{Samuel Williams}, \bibinfo{person}{Mary Hall}, {and}
  \bibinfo{person}{Hans Johansen}.} \bibinfo{year}{2019}\natexlab{}.
\newblock \showarticletitle{Exploiting reuse and vectorization in blocked
  stencil computations on CPUs and GPUs}. In \bibinfo{booktitle}{\emph{SC'19}}.
  \bibinfo{pages}{1--44}.
\newblock


\bibitem[\protect\citeauthoryear{Zumbusch}{Zumbusch}{2012}]%
        {Zumbusch:para12}
\bibfield{author}{\bibinfo{person}{Gerhard Zumbusch}.}
  \bibinfo{year}{2012}\natexlab{}.
\newblock \bibinfo{booktitle}{\emph{Vectorized Higher Order Finite Difference
  Kernels}}.
\newblock \bibinfo{pages}{343--357}.
\newblock
\showISBNx{978-3-642-36803-5}
\urldef\tempurl%
\url{https://doi.org/10.1007/978-3-642-36803-5_25}
\showDOI{\tempurl}


\end{thebibliography}

\end{document}